\documentclass[conference,10pt]{IEEEtran}
\IEEEoverridecommandlockouts

\usepackage{lipsum}
\usepackage{amsmath,amsfonts}
\usepackage{algorithmic}
\usepackage{graphicx}
\usepackage{textcomp}
\usepackage{xcolor}
\usepackage{tikz}
\usepackage{pgfplots}
\pgfplotsset{compat=1.18}
\usepgfplotslibrary{statistics}
\usepackage[frozencache]{minted}
\usepackage{caption}
\usepackage{subcaption}
\usepackage{adjustbox}
\usepackage{soul}
\usepackage{hyperref}
\usepackage{url}

\def\withartifacts{}
\def\withpaper{}

\def\BibTeX{{\rm B\kern-.05em{\sc i\kern-.025em b}\kern-.08em
    T\kern-.1667em\lower.7ex\hbox{E}\kern-.125emX}}
\begin{document}
\newcommand{\bytes}{\mbox{bytes}}
\newcommand{\byte}{\mbox{byte}}
\newcommand{\second}{\mbox{s}}
\newcommand{\seconds}{\mbox{s}}
\newcommand{\flop}{\mbox{flop}}
\newcommand{\flops}{\mbox{flops}}
\newcommand{\NJFLOP}{\mbox{nJ/\flop}}
\newcommand{\instr}{\mbox{instr}}
\newcommand{\cycle}{\mbox{cy}}
\newcommand{\iter}{\mbox{it}}
\newcommand{\cycles}{\mbox{cy}}
\newcommand{\FCY}{\mbox{\flop/\cycle}}
\newcommand{\FIT}{\mbox{\flop/\iter}}
\newcommand{\BIT}{\mbox{\byte/\iter}}
\newcommand{\FR}{\mbox{\flops/\mbox{row}}}
\newcommand{\BR}{\mbox{\byte/\mbox{row}}}
\newcommand{\CYF}{\mbox{\cycles/\flop}}
\newcommand{\CS}{\mbox{\cycle/\second}}
\newcommand{\GCS}{\mbox{G\cycle/\second}}
\newcommand{\word}{\mbox{Word}}
\newcommand{\words}{\mbox{Words}}
\newcommand{\order}[1]{\mbox{${\cal O}\left(\mbox{#1}\right)$}}
\newcommand{\bit}{\mbox{bit}}
\newcommand{\bits}{\mbox{bits}}
\newcommand{\GBPS}{\mbox{G\bit/\second}}
\newcommand{\MBPS}{\mbox{M\bit/\second}}
\newcommand{\FS}{\mbox{\flop/\second}}
\newcommand{\BS}{\mbox{\byte/\second}}
\newcommand{\BC}{\mbox{\byte/\cycle}}
\newcommand{\GBS}{\mbox{G\byte/\second}}
\newcommand{\MBS}{\mbox{M\byte/\second}}
\newcommand{\GWS}{\mbox{G\word/\second}}
\newcommand{\GFS}{\mbox{G\flop/\second}}
\newcommand{\MFS}{\mbox{M\flop/\second}}
\newcommand{\lup}{\mbox{LUP}}
\newcommand{\lups}{\mbox{LUPs}}
\newcommand{\LUPCY}{\mbox{\lup/\cycle}}
\newcommand{\LUPS}{\mbox{\lup/\second}}
\newcommand{\MLUPS}{\mbox{M\lup/\second}}
\newcommand{\GLUPS}{\mbox{G\lup/\second}}
\newcommand{\GHZ}{\mbox{GHz}}
\newcommand{\ns}{\mbox{ns}}
\newcommand{\WF}{\mbox{\word/\flop}}
\newcommand{\BF}{\mbox{\byte/\flop}}
\newcommand{\FB}{\mbox{\flop/\byte}}
\newcommand{\BL}{\mbox{\byte/\lup}}
\newcommand{\GB}{\mbox{GB}}
\newcommand{\KB}{\mbox{kB}}
\newcommand{\MB}{\mbox{MB}}
\newcommand{\GiB}{\mbox{GiB}}
\newcommand{\MiB}{\mbox{MiB}}
\newcommand{\KiB}{\mbox{KiB}}
\newcommand{\TB}{\mbox{TB}}
\newcommand{\TiB}{\mbox{TiB}}
\newcommand{\W}{\mbox{W}}
\newcommand{\muarch}{\mbox{$\mu$-arch}}
\newcommand{\muop}{\mbox{$\mu$-op}}
\newcommand{\muops}{\mbox{$\mu$-ops}}

\newcommand{\eos}{\;.}
\newcommand{\cma}{\;,}
\newcommand{\rlm}{roof{}line model}
\newcommand{\rl}{roof{}line}
\newcommand{\Rlm}{Roof{}line model}
\newcommand{\Rl}{Roof{}line}
\newcommand{\olsep}{\|}
\newcommand{\nolsep}{|}
\newcommand{\ecmspace}{\,}
\newcommand{\TOL}{$T_{\mathrm{c}\_\mathrm{OL}}$}
\newcommand{\NNZR}{$N_\mathrm{nzr}$}
\newcommand{\NR}{$N_\mathrm{r}$}
\newcommand{\NNZ}{$N_\mathrm{nz}$}
\newcommand{\ecm}[6]{\mbox{$\left\{{#1}\ecmspace\olsep\ecmspace {#2}\ecmspace\nolsep\ecmspace {#3}\ecmspace\nolsep\ecmspace {#4}\ecmspace\nolsep\ecmspace {#5}\right\}\ecmspace{#6}$}}
\newcommand{\epsep}{\rceil}
\newcommand{\ecmp}[4]{\mbox{$\left\{{#1}\ecmspace\epsep\ecmspace {#2}\ecmspace\epsep\ecmspace {#3}\right\}\ecmspace{#4}$}}
\newcommand{\ecme}[4]{\mbox{$\left({#1}\ecmspace\epsep\ecmspace {#2}\ecmspace\epsep\ecmspace {#3}\right)\ecmspace{#4}$}}
\newcommand{\sellcs}{SELL-\texorpdfstring{$C$-$\sigma$}{C-sigma}}
\newcommand{\likwid}{\texttt{LIKWID}}
\newcommand{\likwidperfctr}{\texttt{likwid-perfctr}}
\newcommand{\likwidpin}{\texttt{likwid-pin}}
\newcommand{\likwidbench}{\texttt{likwid-bench}}
\newcommand{\lmbench}{\texttt{lmbench}}
\newcommand{\afx}{A64FX}
\newcommand{\spmv}{SpMV}
\newcommand{\cmg}{CMG}
\newcommand{\mve}{MVE}
\newcommand{\crs}{CRS}
\newcommand{\ellpack}{ELLPACK}
\newcommand{\tands}{dRECT}
\newcommand{\bq}{\begin{equation}}
\newcommand{\eq}{\end{equation}}

\newcommand{\clv}{CloverLeaf}

\newcommand{\inputtikz}[2]{%
	\input{#1/#2.tex}%
}

\newenvironment{customlegend}[1][]{%
	\begingroup
	\csname pgfplots@init@cleared@structures\endcsname
	\pgfplotsset{#1}%
}{%
	\csname pgfplots@createlegend\endcsname
	\endgroup
}%
\def\addlegendimage{\csname pgfplots@addlegendimage\endcsname}

\newcommand{\GHacomm}[1]{{\color{red}-- {#1} --\color{black}} }
\newcommand{\JLcomm}[1]{{\color{orange}-- {#1} --\color{black}} }
\newcommand{\DOcomm}[1]{{\color{magenta}-- {#1} --\color{black}} }
\newcommand{\TGcomm}[1]{{\color{tumbleweed}-- {#1} --\color{black}} }
\newcommand{\revon}{\color{black}}
\newcommand{\revoff}{\color{black}}


\makeatletter
\newcommand{\linebreakand}{%
  \end{@IEEEauthorhalign}
  \hfill\mbox{}\par
  \mbox{}\hfill\begin{@IEEEauthorhalign}
}
\makeatother

\definecolor{amber}{rgb}{1.0, 0.75, 0.0}
\definecolor{amethyst}{rgb}{0.6, 0.4, 0.8}
\definecolor{applegreen}{rgb}{0.55, 0.71, 0.0}
\definecolor{tumbleweed}{rgb}{0.87, 0.67, 0.53}
\definecolor{greyback}{RGB}{248,248,248}
\definecolor{deepblue}{RGB}{43,131,186}
\definecolor{greenalt}{RGB}{35,139,69}
\definecolor{darkred}{RGB}{215,25,28}
\definecolor{darkorange}{RGB}{253,174,97}
\definecolor{light-gray}{gray}{0.85}
\definecolor{lighter-gray}{gray}{0.95}
\definecolor{applegreen}{rgb}{0.55, 0.71, 0.0}
\definecolor{asparagus}{rgb}{0.53, 0.66, 0.42}
\definecolor{babyblueeyes}{rgb}{0.63, 0.79, 0.95}
\definecolor{burntsienna}{rgb}{0.91, 0.45, 0.32}
\definecolor{deepcarmine}{rgb}{0.66, 0.13, 0.24}
\definecolor{lightcornflowerblue}{rgb}{0.6, 0.81, 0.93}
\definecolor{steelblue}{rgb}{0.27, 0.51, 0.71}
\definecolor{pastelblue}{rgb}{0.68, 0.78, 0.81}
\definecolor{pastelbrown}{rgb}{0.51, 0.41, 0.33}
\definecolor{pastelgray}{rgb}{0.81, 0.81, 0.77}
\definecolor{pastelgreen}{rgb}{0.47, 0.87, 0.47}
\definecolor{pastelmagenta}{rgb}{0.96, 0.6, 0.76}
\definecolor{pastelorange}{rgb}{1.0, 0.7, 0.28}
\definecolor{pastelpink}{rgb}{1.0, 0.82, 0.86}
\definecolor{pastelpurple}{rgb}{0.7, 0.62, 0.71}
\definecolor{pastelred}{rgb}{1.0, 0.41, 0.38}
\definecolor{pastelviolet}{rgb}{0.8, 0.6, 0.79}
\definecolor{pastelyellow}{rgb}{0.99, 0.99, 0.59}
\definecolor{darkblue}{rgb}{0.0, 0.0, 0.55}

\definecolor{blue-seq-0}{RGB}{255, 255, 204}
\definecolor{blue-seq-1}{RGB}{119, 233, 180}
\definecolor{blue-seq-2}{RGB}{127, 205, 187}
\definecolor{blue-seq-3}{RGB}{ 65, 182, 196}
\definecolor{blue-seq-4}{RGB}{ 44, 127, 184}
\definecolor{blue-seq-5}{RGB}{ 37,  52, 148}
\definecolor{red-seq-0}{RGB}{254, 229, 217}
\definecolor{red-seq-1}{RGB}{252, 187, 161}
\definecolor{red-seq-2}{RGB}{252, 146, 114}
\definecolor{red-seq-3}{RGB}{251, 106, 74}
\definecolor{red-seq-4}{RGB}{222,  45, 38}
\definecolor{red-seq-5}{RGB}{165,  15, 21}
\definecolor{green-seq-0}{RGB}{237,248,233}
\definecolor{green-seq-1}{RGB}{199,233,192}
\definecolor{green-seq-2}{RGB}{161,217,155}
\definecolor{green-seq-3}{RGB}{116,196,118}
\definecolor{green-seq-4}{RGB}{49,163,84}
\definecolor{green-seq-5}{RGB}{0,109,44}
\definecolor{qual-0}{RGB}{102,194,165}
\definecolor{qual-1}{RGB}{252,141,98}
\definecolor{qual-2}{RGB}{141,160,203}
\definecolor{qual-3}{RGB}{231,138,195}
\definecolor{qual-4}{RGB}{166,216,84}
\definecolor{qual-5}{RGB}{255,217,47}
\definecolor{qual-6}{RGB}{229,196,148}
\definecolor{cbf-0}{RGB}{215,48,39}
\definecolor{cbf-1}{RGB}{244,109,67}
\definecolor{cbf-2}{RGB}{253,174,97}
\definecolor{cbf-3}{RGB}{254,224,144}
\definecolor{cbf-4}{RGB}{224,243,248}
\definecolor{cbf-5}{RGB}{171,217,233}
\definecolor{cbf-6}{RGB}{116,173,209}
\definecolor{cbf-7}{RGB}{69,117,180}

\definecolor{longqual-0}{RGB}{166,206,227}
\definecolor{longqual-1}{RGB}{31,120,180}
\definecolor{longqual-2}{RGB}{178,223,138}
\definecolor{longqual-3}{RGB}{51,160,44}
\definecolor{longqual-4}{RGB}{251,154,153}
\definecolor{longqual-5}{RGB}{227,26,28}
\definecolor{longqual-6}{RGB}{253,191,111}
\definecolor{longqual-7}{RGB}{255,127,0}
\definecolor{longqual-8}{RGB}{202,178,214}
\definecolor{longqual-9}{RGB}{106,61,154}
\definecolor{longqual-10}{RGB}{255,255,153}
\definecolor{longqual-11}{RGB}{177,89,40}
\ifdefined\withpaper

\title{CloverLeaf on Intel Multi-Core CPUs: A Case Study in Write-Allocate Evasion}

\author{

\IEEEauthorblockN{
    Jan Laukemann\IEEEauthorrefmark{1},
    Thomas Gruber\IEEEauthorrefmark{1},
    Georg Hager\IEEEauthorrefmark{1}, 
    Dossay Oryspayev\IEEEauthorrefmark{2}, and
    Gerhard Wellein\IEEEauthorrefmark{1} } \linebreakand \\
\IEEEauthorblockA{\IEEEauthorrefmark{1}
    \textit{Friedrich-Alexander-Universität Erlangen-Nürnberg} \\
    \textit{Erlangen National High Performance Computing Center}\\
    Erlangen, Germany\\
    \{jan.laukemann, thomas.gruber, georg.hager, gerhard.wellein\}@fau.de
    } \and \\
\IEEEauthorblockA{\IEEEauthorrefmark{2}
    \textit{Programming Models and Compilers Group} \\
    \textit{Computational Science Initiative} \\
    \textit{Brookhaven National Laboratory} \\
    Upton, NY, USA \\
    doryspaye@bnl.gov
}

}
\IEEEoverridecommandlockouts
\IEEEpubid{\hspace{-1.05\linewidth}\begin{minipage}{\columnwidth}\ \\[32pt] \copyright 2024 IEEE. Personal use of this material is permitted. Permission from IEEE must be
obtained for all other uses, in any current or future media, including
reprinting/republishing this material for advertising or promotional purposes, creating new
collective works, for resale or redistribution to servers or lists, or reuse of any copyrighted
component of this work in other works.\end{minipage}} 

\maketitle
\IEEEpeerreviewmaketitle

\begin{abstract}
In this paper we analyze the MPI-only version of the \clv\ code from the SPEChpc 2021 benchmark suite on recent Intel Xeon ``Ice Lake'' and ``Sapphire Rapids'' server CPUs. We observe peculiar breakdowns in performance when the number of processes is prime. Investigating this effect, we create first-principles data traffic models for each of the stencil-like hotspot loops. With application measurements and microbenchmarks to study memory data traffic behavior, we can connect the breakdowns to SpecI2M, a new write-allocate evasion feature in current Intel CPUs.
For serial and full-node cases we are able to predict the memory data volume analytically with an error of a few percent. We find that if the number of processes is prime, SpecI2M fails to work properly, which we can attribute to short inner loops emerging from the one-dimensional domain decomposition in this case. We can also rule out other possible causes of the prime number effect, such as breaking layer conditions, MPI communication overhead, and load imbalance. 
\end{abstract}


\section{Introduction} \label{sec:intro}
The new SPEChpc 2021 benchmark suite~\cite{spechpc2021} was specifically created for state-of-the-art HPC systems utilizing high parallelism. Its current version~1.1 was released in July 2022.
It tries to address challenges of real-world applications with different sizes of workloads and to provide comparative performance metrics for both CPU and GPU runs with support for OpenACC, OpenMP, and MPI.
\clv~\cite{cloverleaf, Mallinson2013CloverLeafPH}, a Lagrangian-Eulerian hydrodynamics mini-app developed as part of the Mantevo project~\cite{mantevo}, is part of the suite.
In this work we conduct a performance study of the pure MPI version of the \clv\ benchmark on the Intel Ice Lake SP~(ICX) server hardware platform. Guided by intriguing performance effects that we ascribe to a newly introduced write-allocate evasion feature \revon (``SpecI2M'')\revoff, we develop a memory traffic model for all 22 loops in the three most time-consuming functions in the code (representing 69\% of the overall runtime) and validate this model against measurements.
\revon We further investigate the SpecI2M feature in more detail on both the ICX and Intel Sapphire Rapids~(SPR) architecture and conduct several experiments to study its behavior and to identify conditions under which the hardware feature works as intended and where it fails. 
Finally, we show that write-allocate evasion works best if large arrays are written consecutively and that the \clv\ code gives best results when non-temporal stores are employed on top and, thus, provide recommendations for optimizing the benchmark. \revoff 

This paper is organized as follows: In Section~\ref{sec:background}, we cover related work as well as the basic theory for performance modeling.
Section~\ref{sec:methodology} details the hardware and software testbed used for the experiments.
In Section~\ref{sec:results}, we first show profiling and modeling of the \clv\ mini-app, after which we focus on its odd behavior when using a prime number of cores. 
Finally, Section~\ref{sec:conclusion} summarizes the work and discusses future work.

\section{Background and Related Work} \label{sec:background}

\subsection{\Rlm}\label{sec:RLM}

Analytic performance models yield valuable insight into hardware bottlenecks and opportunities for optimization. 
The most popular analytic performance model is the \Rlm~\cite{hockney89,roofline:2009}.
It combines a machine model with an application model to give upper performance limits for individual loops in a program. 
 The machine model comprises of, in the simplest case, two performance bottlenecks: the maximum in-core performance $P_\mathrm{max}$ and the memory bandwidth $b_S$. The application model is given per loop, as the ratio  $I$ of work (e.g., in \flops) to memory transfer data volume limit (in \byte), which is termed \emph{computational intensity}. An upper limit for a loop with intensity $I$ is thus $P=\min\left(P_\mathrm{max},I\times b_S\right)$. The graph of this function, when plotted with respect to $I$, resembles the shape of a roof when plotted; hence the name ``\Rlm.'' Although this model is crude and many refinements are possible~\cite{Hager2012,RLLoft:2013,Stengel2015}, it is invaluable for initial performance assessment and analysis. In practice, an actual \Rl\ plot is often unnecessary since it is more efficient to just work with the numbers.

For memory-bound code such as the loops present in \clv, only the $I\times b_S$ limit is relevant. The \emph{code balance} $B_c=I^{-1}$ is often used instead of the intensity because its value is usually an integer number that is easily interpretable, especially when a loop iteration (``$\iter$'') instead of a \flop\ is used as the unit of work. The \Rl\ limit is then $b_S/B_c$, and the only viable optimization approach is to lower the code balance by reducing the memory data transfer volume. 

Note that in cases where the memory bandwidth $b_S$ can be fully saturated, establishing a model for memory-bound loops boils down to an accurate prediction of the memory data volume.

\subsection{SPEChpc and \clv}

The Standard Performance Evaluation Corporation (SPEC)\footnote{\url{https://www.spec.org}} was founded in 1988 to develop standardized benchmarks and tools for evaluating performance and other key metrics on state-of-the-art computing systems.
The subgroup SPEC High-Performance Group (HPG) publishes benchmark suites specifically for MPI, OpenMP, and accelerators.
In 2021, a general HPC suite SPEChpc 2021 covering many aspects of HPC applications on heterogeneous compute nodes was released.

The \clv~benchmark~\cite{cloverleaf, Mallinson2013CloverLeafPH} is a Lagrangian-Eulerian Hydrodynamics mini-app. It is part of the SPEChpc2021 suite~\cite{spechpc2021} but originates from the Mantevo benchmark suite~\cite{mantevo}.
The SPEChpc2021 version combines MPI, OpenMP and Open\-ACC into one code.
The code solves the compressible Euler equations on a Cartesian 2D grid with an explicit second-order accurate method working on the grid cells.
It uses a staggered grid with information stored in the grid cell center as well as vectors stored at the cell corners.
The code is split into multiple kernels, each applying vector or stencil computations on the grid.
\revon For communicating domain boundaries, halos of a size of 2--5 (depending on the kernel) are used. \revoff

\subsection{Stencils and layer conditions}

In stencil computations, each element is calculated by applying approximations including a fixed pattern of surrounding elements.
In case of the simple regular stencils used in \clv, \emph{layer conditions} determine the required data traffic from memory per grid point update. Layer conditions are an adaptation of the reuse distance concept to stencil algorithms~\cite{Stengel2015}.
Depending on the extent of the stencil in the outer grid dimension, multiple rows (``layers'') of the grid need to be cached in order to re-use as much data as possible for minimum code balance (\BF) and thus maximum performance. 
\begin{figure}[tb]\centering
\includegraphics*[width=0.9\linewidth]{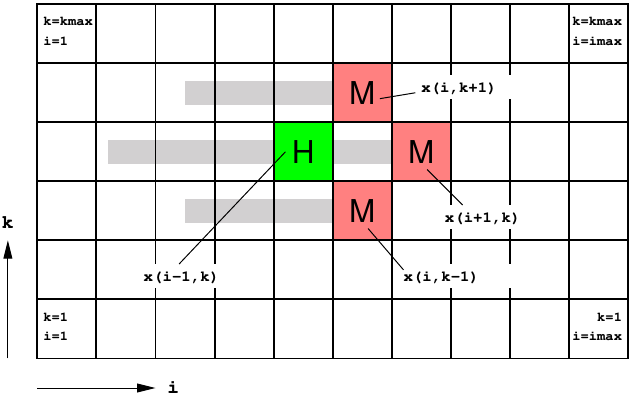}\medskip\\
\includegraphics*[width=0.9\linewidth]{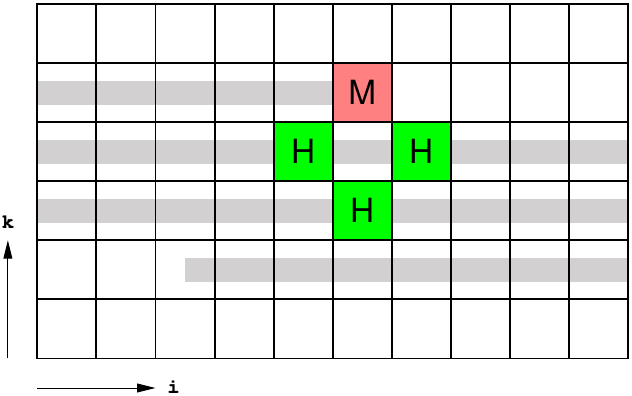}
\caption{Visualization of layer conditions (LCs) with a 2D 4-point stencil (right-hand side shown only). Shaded areas are cached elements. \emph{Top}: LC broken, the cache is not large enough to hold three successive grid rows. Three out of four accesses are cache misses. \emph{Bottom}: LC satisfied, the cache can fit at least three rows. Only one of four accesses is a miss.\label{fig:LC}}
\end{figure}
\begin{listing}
\begin{minted}[
frame=lines,
framesep=2mm,
baselinestretch=1.2,
bgcolor=lighter-gray,
fontsize=\footnotesize,
]{fortran}
do k=2,kmax-1
  do i=2,imax-1
    y(i,k) = 0.25d0 * (      x(i,k+1) + &
                        x(i-1,k) + x(i+1,k) + &
                             x(i,k-1) )
  enddo
enddo
\end{minted}
\caption{Sample Fortran code of a 2D 4-point stencil with radius 1. The arrangement of the right-hand side neighbors corresponds to  Fig.~\ref{fig:LC}.}
\label{lst:f90_stencil}
\vspace{-5pt}
\end{listing}

Figure~\ref{fig:LC} and Listing~\ref{lst:f90_stencil} illustrate this concept on the example of a 2D 4-point stencil. The stencil pattern shows which elements of the array \verb.x(:,:). are required per loop iteration. If the cache is small (\emph{Top} of Figure~\ref{fig:LC}), i.e., if it cannot hold three successive rows of the grid, elements brought into the cache by the three rows of the stencil do not stay cached long enough to be picked up in the sweep over the next grid row, causing three elements (labeled ``M'') to be loaded from memory or a lower cache level. However, if the cache is large enough (\emph{Bottom} of Figure~\ref{fig:LC}), two of the three misses can come from cache and only \verb.x(i,k+1). must be fetched from far away. The layer condition thus reads as
\bq\label{eq:LC}
3\times\text{\texttt{imax}}\times\text{\texttt{sizeof(double)}} < C
\eq
in this case, where $C$ is the available cache size.\footnote{In practice, $C$ is often chosen to be half the actual cache size, but in cases where more arrays are accessed concurrently, it could be less. One approach is to assume that half the cache is shared by all arrays that are read or written.} If it is fulfilled for the L3 cache, the code balance in memory is $24\,\BIT=6\,\BF$; if it is broken, the code balance grows to $40\,\BIT=10\,\BF$. 

\revon Typical loop optimizations such as loop tiling aim at re-establishing a layer condition by introducing an additional loop on the outside of the nest, which iterates over blocks of the inner loop(s)~\cite{Hager2010}. The block size is then a free parameter which takes the role of \verb.imax. in (\ref{eq:LC}) and can be chosen so that a layer condition is fulfilled. 

If the hardware supports it, pure ``streaming'' arrays, i.e., with no temporal reuse within a sweep, could be restricted to using just a small section of the cache in order to leave more room for the layer condition. This \emph{sector cache} feature is, e.g., available on the Fujitsu A64FX CPU~\cite{Alappat2022} but not on current x86 designs.\revoff

Of course, the condition depends on the shape of the stencil. For instance, if it accesses only two rows in the outer dimension, only two rows need to fit and the first factor in Equation~\ref{eq:LC} becomes $2$. If the LC applies to a shared cache, $C$ is the available cache size per process or thread.

\subsection{Write-allocates and their evasion}

Modern processors make heavy use of caches to boost performance by keeping recently used data close to the execution units. In general, CPU cores can only read or write data that is in the L1 cache. This means that in case of a  (read or write) cache miss, the corresponding cache line must be fetched from a lower cache level or from main memory first.
A transfer triggered by a write miss is called a \emph{write-allocate} (WA) or \emph{read-for-ownership} (RFO). 
At some point after the actual write operation to the L1 cache is finished, the cache line is evicted back to a lower level and eventually to main memory. 
Of course, if a core overwrites the whole cache line anyway, the WA transfer is basically unnecessary.

Many CPU architectures provide mechanisms to avoid these transfers; one option are special store instructions with \emph{non-temporal hints}. The x86 and ARM architectures offer non-temporal (NT) store instructions (a.k.a.\ \emph{streaming stores}), which bypass the normal cache hierarchy and write directly to memory.\footnote{This is a simplification; NT stores actually write to a small write-combine buffer which is later flushed to memory.}

In some ARM-based CPUs such as the Fujitsu A64FX~\cite{Alappat2022,A64FXmanual}, a special \emph{cache line claim} (or \emph{cache line zero}) instruction claims a cache line in the cache without fetching it from memory. Of course, such instructions must be used with care in a multi-core environment to avoid coherence issues. An important difference to non-temporal stores is that cache line zero allows immediate reuse of the written data from the cache, but it also takes up space that could be used otherwise if this is not required.

Finally, cache line claim can be automated by the hardware without the need for special instructions. If the hardware detects that a cache line will be overwritten entirely (e.g., by monitoring entries in the store buffers between the registers and the L1 cache), the line can be claimed in the cache right away. This has been available for some time on ARM-based CPUs such as the
Neoverse N1~\cite{NeoverseTM}, and it is now also featured by Intel Ice Lake SP and Sapphire Rapids architectures. In the following we will call this mechanism \emph{write-allocate evasion}. Although limited general information is published by the vendors, the detailed mechanisms of WA evasion are not disclosed.

\section{Methodology} \label{sec:methodology}
Experiments were conducted on a two-socket Intel Xeon Platinum 8360Y ``Ice Lake SP'' node with 36 cores and 128\,\GiB\ of DDR4-3200 memory per socket. Each core has private 48\,\KiB ~L1 and 1.25\,\MiB ~L2 caches; the 54\,\MiB ~L3 cache is shared across a socket. 
Transparent Huge Pages~(THP) and NUMA balancing were activated and Sub-NUMA Clustering (SNC) was turned on, resulting in two NUMA domains per socket.
The core clock frequency was always fixed to 2.4\,\GHZ.
For the experiments on the ``Sapphire Rapids'' platform, we used \revon two different systems: (i) a two-socket Intel Xeon Pla\-ti\-num 8470 node with 52 cores and 512\,\GiB\ memory and (ii)\revoff\ a two-socket Intel Xeon Platinum 8480+ node with 56 cores and
\revon\ 256\,\GiB\ memory per socket. Both servers use 8 channels of DDR5-4800 RAM per socket 
and are equipped with the same ``Golden Cove'' cores, each with\revoff\ a private 48\,\KiB\ L1 and 2.0\,\MiB\ L2 cache; the 105\,\MiB\ L3 cache is shared across a socket. While THP and NUMA balancing is\revon\ always\revoff\ activated, SNC is\revon\ adjustable and set as described on the 8470 node and always off on the 8480+ node,\revoff\ i.e., here each socket represents a single NUMA domain.
The core clock frequency was fixed to its base frequency of 2.0\,\GHZ\revon\ for both systems. \revoff

The servers ran AlmaLinux~8.8 and all code was compiled with the Intel legacy ICC compiler v2021.6.0 and Intel MPI v2021.7.0.
For hardware performance counter measurements \revon such as memory bandwidth, read and write data volume, and FLOPs,\revoff\ we use \verb.likwid-perfctr. from the LIKWID tool suite in version 5.2.2\footnote{For Sapphire Rapids support, the at the time of writing unreleased pull request \url{https://github.com/RRZE-HPC/likwid/pull/524} was used.}~\cite{likwid}.
For all runs we used the unchanged input file of the ``Tiny'' working set provided by SPEChpc2021~(termed ``\texttt{519.clvleaf\_t}''), i.e., running 400 iterations on a square grid of size 15,360$^2$, in MPI-only mode.
Furthermore, we adjusted the code to allow for more flexibility:
\begin{itemize}
    \item align the arrays used for computation to 64\,\bytes,
    \item mark each loop inside the hotspot functions with region markers for hardware counter measurements,
    \item add MPI barriers before and after each loop inside the hotspot functions,
    \item apply non-temporal store directives (``\texttt{!DIR\$ vector nontemporal()}'') to each inner loop in the hotspot functions.
\end{itemize}
These knobs allow us to measure exactly the data volume of each loop and easily turn code optimizations on and off.
Since entering and leaving the marked regions for hardware counter measurements and the introduction of barriers obviously increase the runtime of the application, those options were only used for measuring non-time-related hardware events, most importantly read and write memory data volume from and to main memory.

\section{Experiments and results} \label{sec:results}

All experiments and results for reproducing the findings of this work can be found at\revon\ \url{https://zenodo.org/doi/10.5281/zenodo.8407657}.\revoff
The artifact contains a detailed description for each data set and Figure used in this paper and furthermore provides a link to download all results created for this work 
from a ``Zenodo'' repository.

\subsection{Performance profiling and modeling of \clv} \label{ssec:modeling}

\begin{figure*}[tp]
	\centering
	%
	\begin{tikzpicture}
    \centering
    \begin{axis}[
        height=4.5cm,
        width=0.95*\linewidth,
        ymajorgrids=true,
        tick align=inside,
        axis y line*=right,
        axis x line=none,
        ylabel={Mem Bandwidth [GB/s]},
        ymin=0,
        ymax=400,        
        xmin=0,
        xmax=72,
        ytick = {20, 100, 200, 300, 400},
        xtick = {},
    ]
    \addplot[draw=asparagus, line width=.4mm] table[x expr=\thisrowno{0}, y expr=\thisrowno{9}/1000, header=true, col sep=comma] {data/clv_naive_memdp.csv}; \label{plot:bw}
    \end{axis}
    
    \begin{axis}[
        height=4.5cm,
        width=0.95*\linewidth,
        ymajorgrids=true,
        tick align=inside,
        xlabel={Ranks},
        ylabel={Speedup},
        ymin=0,
        ymax=20,        
        xmin=0,
        xmax=72,
        ytick = {1, 5, 10, 15, 20},
        xtick = {1,3,5,7,11,15,17,19,23,25,29,31,35,37,41,43,45,47,53,55,59,61,65,67,71},
        legend pos=north west,
        legend style={draw=none, legend columns=1, at={(0.01, 0.9)}},
        label style={font=\small},
        tick label style={font=\small},
    ]
    \draw[dotted, draw=gray, line width=.5mm,] (18, 0) -- (18, 20);
    \addplot[draw=pastelorange, line width=.4mm] table[x expr=\thisrowno{0}, y expr=\thisrowno{3}, header=true, col sep=comma] {data/scaling.csv};
    \label{speedup}
    \addlegendentry{Speedup}
    \addlegendimage{/pgfplots/refstyle=plot:bw}\addlegendentry{Bandwidth}
    \end{axis}

\end{tikzpicture}%

	\caption{Speedup of the \clv ~mini-app versus number of MPI processes on an Intel Ice Lake SP server with compact pinning. The dotted gray line marks the end of the first ccNUMA domain. The data points represent the median out of ten separate runs with error bars omitted due to variations being negligible (maximum deviation of 2.5\,\%).}
	\label{fig:speedup}
\end{figure*}
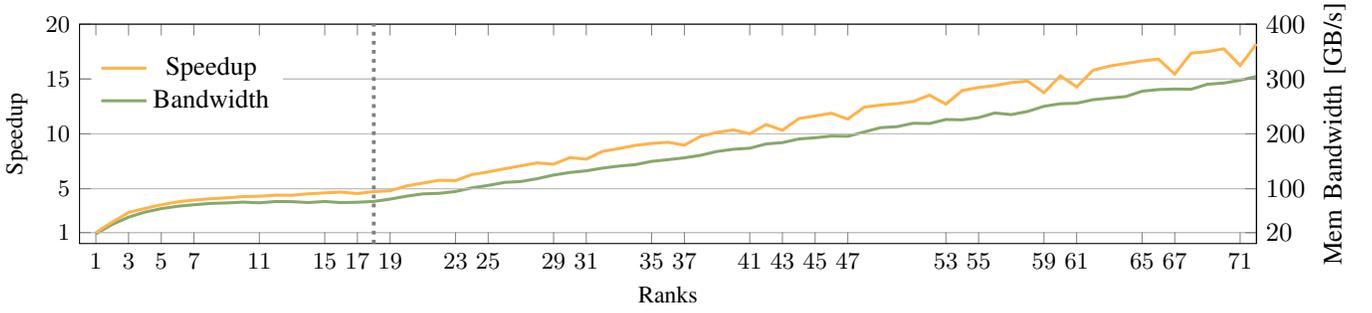
To get an initial idea of the performance features of the mini-app, we execute a simple scaling run of the original version of \clv\ with compact pinning (see Fig.~\ref{fig:speedup})\revon, i.e., filling each ccNUMA domain and each socket first before assigning work to hardware threads of the next domain,\revoff\ and measure the memory bandwidth using hardware performance counters at memory controller level (LIKWID's \texttt{MEM} performance group).
One can clearly identify a memory bandwidth saturation pattern on the first ccNUMA domain (18 cores) with the attainable peak memory bandwidth being reached at about 9 cores.
\revon This---as well as the bandwidth behavior for a core count exceeding one ccNUMA domain---matches the overall achievable memory bandwidth on this system one can obtain with any memory-bound streaming code such as the STREAM bandwidth~\cite{McCalpin1995}.\revoff However, comparing the speedup with the actual (measured) memory bandwidth, one can see that only the bandwidth saturates while the speedup continues to rise even beyond saturation.

Furthermore, the expected linear scaling across the rest of the node is disturbed by performance drops where the number of processes is prime. It is remarkable that these drops in speedup are not accompanied by drops in memory bandwidth. 
We therefore apply the \texttt{gprofng} profiler provided in the GNU Binutils collection~\cite{binutils} to identify the hotspot functions of the code, as shown in Listing~\ref{lst:gprofng}.
\begin{listing}
\begin{minted}[
frame=lines,
framesep=2mm,
baselinestretch=1.2,
bgcolor=lighter-gray,
fontsize=\footnotesize,
]{text}
Name                           Excl. Total  CPU
                                  sec.        %
 <Total>                       36344.824  100.00
 advec_mom_kernel              12998.162   35.76
 advec_cell_kernel              7560.869   20.80
 pdv_kernel                     4553.785   12.53
 accelerate_kernel              1953.466    5.37
 ideal_gas_kernel               1894.885    5.21
 flux_calc_kernel               1649.954    4.54
 reset_field_kernel             1599.509    4.40
 calc_dt_kernel                 1210.607    3.33
 viscosity_kernel                920.064    2.53
\end{minted}
\caption{\texttt{gprofng} profile of a 72-rank run of \clv, showing the exclusive runtime of each function. The output was limited to the ten most time-consuming functions.}
\label{lst:gprofng}
\vspace{-5pt}
\end{listing}

The three functions \texttt{advec\_mom\_kernel}, \texttt{ad\-vec\_cell\_ker\-nel}, and \texttt{pdv\_kernel} contributes 69\,\% of the overall runtime.
While Listing~\ref{lst:gprofng} only shows the profile of a 72-rank run, the aggregate runtime of these hotspots stays consistently between 67.5\,\% and 69.2\,\% throughout any number of ranks between 1 and 72.
The \texttt{advec\_mom} function consists of twelve loops, \texttt{advec\_cell} of eight, and the \texttt{pdv} comprises two loops.
In order to understand the performance features of these loops, we calculate the memory data transfer volume and the number of \flops\ for each loop as shown in Table~\ref{tab:clv-analysis}.
To this end, we count the number of elements to be read and differentiate here between the minimum and maximum due to layer conditions~(LCs) that can be fulfilled ($\rightarrow$ LCF) or broken ($\rightarrow$ LCB).
Furthermore, we count the number of elements written per loop iteration and how many of those must be read in beforehand, i.e., are an update of the value, or could be written without any previous write-allocates~(WAs).
With these metrics, we can create several upper and a lower limits of the code balance:
\begin{itemize}
\item[(i)] The minimum number (``min'') of \BIT, i.e., with fulfilled LC and no WAs,
\item[(ii)] fulfilled LC and WAs (``LCF,WA''),
\item[(iii)] broken LC and no WAs (``LCB''),
\item[(iv)] broken LC and WAs (``max'')
\end{itemize}
We can compare these numbers with the actual measured code balance of a single-core run ($\BIT_\textrm{meas,1}$).
We already know that the code is memory bound, so the upper performance limit as given by the \Rlm\ (see Sec.~\ref{sec:RLM}) is obtained by dividing the memory bandwidth by the code balance.
\begin{table*}[tp]
    \centering
    \footnotesize
    \begin{tabular}{c|c|c|c|c|c|c|c|c|c|c|c}        
        loop &\#arrays & $\textrm{RD}_{\textrm{LCF}}$/it & $\textrm{RD}_{\textrm{LCB}}$/it & WR/it & (RD\&WR)/it & \flops/it & $\textrm{\BIT}_{\textrm{min}}$ & $\textrm{\BIT}_{\textrm{LCF,WA}}$ & $\textrm{\BIT}_{\textrm{LCB}}$ & $\textrm{\BIT}_{\textrm{max}}$ & $\textrm{\BIT}_{\textrm{meas,1}}$ \\ 
        \hline                                                                
        am00  & 5  & 3  & 4  & 2 & 0 & 4  & 40  & 56 & 48 & 64 & 56.32 \\     
        am01  & 5  & 3  & 4  & 2 & 0 & 4  & 40  & 56 & 48 & 64 & 56.28 \\     
        am02  & 4  & 2  & 3  & 2 & 0 & 2  & 32  & 48 & 40 & 56 & 48.25 \\     
        am03  & 4  & 2  & 2  & 2 & 0 & 2  & 32  & 48 & 32 & 48 & 48.15 \\     
        am04  & 2  & 1  & 2  & 1 & 0 & 4  & 16  & 24 & 24 & 32 & 24.05 \\     
        am05  & 5  & 3  & 5  & 2 & 0 & 10 & 40  & 56 & 56 & 72 & 56.97 \\     
        am06  & 4  & 3  & 3  & 1 & 0 & 9  & 32  & 40 & 32 & 40 & 40.22 \\     
        am07  & 4  & 4  & 4  & 1 & 1 & 4  & 40  & 40 & 40 & 40 & 40.08 \\     
        am08  & 2  & 1  & 2  & 1 & 0 & 4  & 16  & 24 & 24 & 32 & 24.06 \\     
        am09  & 5  & 3  & 6  & 2 & 0 & 10 & 40  & 56 & 64 & 80 & 56.56 \\     
        am10  & 4  & 3  & 5  & 1 & 0 & 8  & 32  & 40 & 48 & 56 & 41.49 \\     
        am11  & 4  & 4  & 5  & 1 & 1 & 4  & 40  & 40 & 48 & 48 & 40.08 \\     
        ac00  & 5  & 3  & 4  & 2 & 0 & 6  & 40  & 56 & 48 & 64 & 56.33 \\     
        ac01  & 4  & 2  & 2  & 2 & 0 & 2  & 32  & 48 & 32 & 48 & 48.25 \\     
        ac02  & 6  & 4  & 4  & 2 & 0 & 17 & 48  & 64 & 48 & 64 & 64.70 \\     
        ac03  & 6  & 6  & 6  & 2 & 2 & 10 & 64  & 64 & 64 & 64 & 64.45 \\     
        ac04  & 5  & 3  & 4  & 2 & 0 & 6  & 40  & 56 & 48 & 64 & 56.29 \\     
        ac05  & 4  & 2  & 3  & 2 & 0 & 2  & 32  & 48 & 40 & 56 & 48.33 \\     
        ac06  & 6  & 4  & 8  & 2 & 0 & 17 & 48  & 64 & 80 & 96 & 66.24 \\     
        ac07  & 6  & 6  & 9  & 2 & 2 & 10 & 64  & 64 & 88 & 88 & 64.85 \\     
        pdv00 & 11 & 9  & 12 & 2 & 0 & 49 & 88  &104 &112 &128 &104.73 \\     
        pdv01 & 13 & 11 & 16 & 2 & 0 & 45 & 104 &120 &144 &160 &120.77 \\     
    \end{tabular}
    \caption{\clv\ performance model input for each loop in the hotspot functions. ``\#arrays'' describes the number of arrays accessed, ``$\textrm{RD}_{\textrm{LCF}}$'' is the number of elements to be read when the layer condition~(LC) is fulfilled and ``$\textrm{RD}_{\textrm{LCB}}$'' is the maximum number of elements to be read due to broken LCs. ``WR'' represents the number of elements to be written, and ``RD\&WR'' is the number of elements out of ``WR'' that are also explicitly read before, while ``\flops/it'' is the number of \flops\ per iteration. The ``$\textrm{\BIT}$'' values derive from the previous columns and can be differentiated into four cases as described in the text. Finally, the last column shows the measured code balance for a single-core run. All loops are labeled (first column) in order of appearance in the function with ``am'', ``ac'', and ``pdv'' being abbreviations for the \texttt{advec\_mom}, \texttt{advec\_cell}, and \texttt{pdv} kernel, respectively.}
    \label{tab:clv-analysis}
\end{table*}
\begin{listing}
\begin{minted}[
frame=lines,
framesep=2mm,
baselinestretch=1.2,
bgcolor=lighter-gray,
fontsize=\footnotesize,
]{fortran}
DO k=y_min,y_max+1
  DO j=x_min-2,x_max+2
    node_flux(j,k)=0.25_8*( mass_flux_x(j  ,k-1) &
                           +mass_flux_x(j  ,  k) &
                           +mass_flux_x(j+1,k-1) &
                           +mass_flux_x(j+1,  k))
  ENDDO
ENDDO
\end{minted}
\caption{Loop nest ``am04'' from the \texttt{advec\_mom} module. For the minimal code balance of 16\,\BIT, the entries {\tt (j,k-1)}, {\tt (j,k)} and {\tt (j+1,k-1)} of \texttt{mass\_flux\_x(:,:)} need to be present in the cache, and \texttt{mass\_flux\_x(j+1,k)} is loaded from memory. Moreover, the write-allocate transfers to \texttt{node\_flux(:,:)} must be evaded.}
\label{lst:advec_mom_loop_04}
\vspace{-5pt}
\end{listing}

Listing~\ref{lst:advec_mom_loop_04} shows the loop nest ``am04'' as an example. Here, the 8\,\bytes\ for \verb.mass_flux_x(j+1,k). always come from memory. The layer condition requires two rows of \verb.mass_flux_x(:,:). to fit in the cache. If it is broken, this array causes 16\,\bytes\ of memory traffic per iteration instead because \verb.mass_flux_x(j+1,k-1). comes from memory, too (\verb.mass_flux_x(j,:). will always come from cache). If the write-allocate transfers to the target array \verb.node_flux(:,:). cannot be avoided, another 16\,\bytes\  are added on top. If some WA mechanism can be leveraged, this reduces to 8\,\bytes. Hence, the minimum code balance for this loop is 16\,\BIT, and the maximum is 32\,\BIT. If \emph{either} the LC is broken \emph{or} WA evasion is impossible, the code balance is 24\,\BIT.


\subsection{Loop categorization} 
Comparing the measured code balance in column $\textrm{\BIT}_{\textrm{meas,1}}$ of Table~\ref{tab:clv-analysis} with the different limits of our model, we can see that the numbers match the predictions for fulfilled LCs and WA.
This meets the expectations since by default the code does not use any non-temporal stores and we never need to fit more than two rows of the grid in the cache to fulfill the LC; i.e., with a maximum row size of $M = 15360$ a lower limit for the required cache size \(C\) is obtained from the condition
\bq
2 \times M \times 8\,\bytes < C/2~,
\eq
which leads to $C > 492\,\KB$. 
The measured code balance for all process counts and kernels can be found in Figure~\ref{fig:code_balance_naive}.

\begin{figure*}[tp]
    \centering
    \begin{subfigure}[b]{0.47\textwidth}
        \begin{adjustbox}{width=\linewidth}
	%
	\begin{tikzpicture}
    \centering
    \begin{axis}[
        height=7cm,
        width=1.1\linewidth,
        ymajorgrids=true,
        minor tick num=4,
        xlabel={Ranks},
        ylabel={Bytes/it},
        ymin=10,
        ymax=60,        
        xmin=0,
        xmax=72,
        legend pos=north east,
        legend style={draw=none, legend columns=1, at={(1.26, 1.0)}},
        label style={font=\small},
        tick label style={font=\small},
    ]
    \draw[dotted, draw=gray, line width=.5mm] (18, 0) -- (18, 130);

    \addplot[draw=pastelblue, line width=.4mm] table[x expr=\thisrowno{0}, y expr=\thisrowno{5}*1000*1000*1000/\thisrowno{2}/15360/15360, header=true, col sep=comma] {data/clv_aligned_hotspots_barriers_memdp.csv};
    \addlegendentry{am00}
    \addplot[draw=pastelbrown, line width=.4mm] table[x expr=\thisrowno{0}, y expr=\thisrowno{11}*1000*1000*1000/\thisrowno{8}/15360/15360, header=true, col sep=comma] {data/clv_aligned_hotspots_barriers_memdp.csv};
    \addlegendentry{am01}
    \addplot[draw=pastelgray, line width=.4mm] table[x expr=\thisrowno{0}, y expr=\thisrowno{17}*1000*1000*1000/\thisrowno{14}/15360/15360, header=true, col sep=comma] {data/clv_aligned_hotspots_barriers_memdp.csv};
    \addlegendentry{am02}
    \addplot[draw=pastelgreen, line width=.4mm] table[x expr=\thisrowno{0}, y expr=\thisrowno{23}*1000*1000*1000/\thisrowno{20}/15360/15360, header=true, col sep=comma] {data/clv_aligned_hotspots_barriers_memdp.csv};
    \addlegendentry{am03}
    \addplot[draw=pastelmagenta, line width=.4mm] table[x expr=\thisrowno{0}, y expr=\thisrowno{29}*1000*1000*1000/\thisrowno{26}/15360/15360, header=true, col sep=comma] {data/clv_aligned_hotspots_barriers_memdp.csv};
    \addlegendentry{am04}
    \addplot[draw=pastelorange, line width=.4mm] table[x expr=\thisrowno{0}, y expr=\thisrowno{35}*1000*1000*1000/\thisrowno{32}/15360/15360, header=true, col sep=comma] {data/clv_aligned_hotspots_barriers_memdp.csv};
    \addlegendentry{am05}
    \addplot[draw=pastelpink, line width=.4mm] table[x expr=\thisrowno{0}, y expr=\thisrowno{41}*1000*1000*1000/\thisrowno{38}/15360/15360, header=true, col sep=comma] {data/clv_aligned_hotspots_barriers_memdp.csv};
    \addlegendentry{am06}
    \addplot[draw=pastelpurple, line width=.4mm] table[x expr=\thisrowno{0}, y expr=\thisrowno{47}*1000*1000*1000/\thisrowno{44}/15360/15360, header=true, col sep=comma] {data/clv_aligned_hotspots_barriers_memdp.csv};
    \addlegendentry{am07}
    \addplot[draw=pastelred, line width=.4mm] table[x expr=\thisrowno{0}, y expr=\thisrowno{53}*1000*1000*1000/\thisrowno{50}/15360/15360, header=true, col sep=comma] {data/clv_aligned_hotspots_barriers_memdp.csv};
    \addlegendentry{am08}
    \addplot[draw=pastelviolet, line width=.4mm] table[x expr=\thisrowno{0}, y expr=\thisrowno{59}*1000*1000*1000/\thisrowno{56}/15360/15360, header=true, col sep=comma] {data/clv_aligned_hotspots_barriers_memdp.csv};
    \addlegendentry{am09}
    \addplot[draw=pastelyellow, line width=.4mm] table[x expr=\thisrowno{0}, y expr=\thisrowno{65}*1000*1000*1000/\thisrowno{62}/15360/15360, header=true, col sep=comma] {data/clv_aligned_hotspots_barriers_memdp.csv};
    \addlegendentry{am10}
    \addplot[draw=darkblue, line width=.4mm] table[x expr=\thisrowno{0}, y expr=\thisrowno{71}*1000*1000*1000/\thisrowno{68}/15360/15360, header=true, col sep=comma] {data/clv_aligned_hotspots_barriers_memdp.csv};
    \addlegendentry{am11}
    
    \end{axis}

\end{tikzpicture}%

        \end{adjustbox}
    \end{subfigure}
    \hspace{0.03\textwidth}
    \begin{subfigure}[b]{0.47\textwidth}
        \begin{adjustbox}{width=\linewidth}
	%
	\begin{tikzpicture}
    \centering
    \begin{axis}[
        height=7cm,
        width=1.1\linewidth,
        ymajorgrids=true,
        minor tick num=4,
        xlabel={Ranks},
        ylabel={Bytes/it},
        ymin=40,
        ymax=130,        
        xmin=0,
        xmax=72,
        legend pos=north east,
        legend style={draw=none, legend columns=1, at={(1.26, 1.0)}},
        label style={font=\small},
        tick label style={font=\small},
    ]
    \draw[dotted, draw=gray, line width=.5mm,] (18, 0) -- (18, 130);
    
    \addplot[draw=pastelblue, line width=.4mm] table[x expr=\thisrowno{0}, y expr=\thisrowno{77}*1000*1000*1000/\thisrowno{74}/15360/15360, header=true, col sep=comma] {data/clv_aligned_hotspots_barriers_memdp.csv};
    \addlegendentry{ac00}
    \addplot[draw=pastelbrown, line width=.4mm] table[x expr=\thisrowno{0}, y expr=\thisrowno{83}*1000*1000*1000/\thisrowno{80}/15360/15360, header=true, col sep=comma] {data/clv_aligned_hotspots_barriers_memdp.csv};
    \addlegendentry{ac01}
    \addplot[draw=pastelgray, line width=.4mm] table[x expr=\thisrowno{0}, y expr=\thisrowno{89}*1000*1000*1000/\thisrowno{86}/15360/15360, header=true, col sep=comma] {data/clv_aligned_hotspots_barriers_memdp.csv};
    \addlegendentry{ac02}
    \addplot[draw=pastelgreen, line width=.4mm] table[x expr=\thisrowno{0}, y expr=\thisrowno{95}*1000*1000*1000/\thisrowno{92}/15360/15360, header=true, col sep=comma] {data/clv_aligned_hotspots_barriers_memdp.csv};
    \addlegendentry{ac03}
    \addplot[draw=pastelmagenta, line width=.4mm] table[x expr=\thisrowno{0}, y expr=\thisrowno{101}*1000*1000*1000/\thisrowno{98}/15360/15360, header=true, col sep=comma] {data/clv_aligned_hotspots_barriers_memdp.csv};
    \addlegendentry{ac04}
    \addplot[draw=pastelorange, line width=.4mm] table[x expr=\thisrowno{0}, y expr=\thisrowno{107}*1000*1000*1000/\thisrowno{104}/15360/15360, header=true, col sep=comma] {data/clv_aligned_hotspots_barriers_memdp.csv};
    \addlegendentry{ac05}
    \addplot[draw=pastelpink, line width=.4mm] table[x expr=\thisrowno{0}, y expr=\thisrowno{113}*1000*1000*1000/\thisrowno{110}/15360/15360, header=true, col sep=comma] {data/clv_aligned_hotspots_barriers_memdp.csv};
    \addlegendentry{ac06}
    \addplot[draw=pastelpurple, line width=.4mm] table[x expr=\thisrowno{0}, y expr=\thisrowno{119}*1000*1000*1000/\thisrowno{116}/15360/15360, header=true, col sep=comma] {data/clv_aligned_hotspots_barriers_memdp.csv};
    \addlegendentry{ac07}
    \addplot[draw=pastelred, line width=.4mm] table[x expr=\thisrowno{0}, y expr=\thisrowno{125}*1000*1000*1000/\thisrowno{122}/15360/15360, header=true, col sep=comma] {data/clv_aligned_hotspots_barriers_memdp.csv};
    \addlegendentry{pdv00}
    \addplot[draw=pastelviolet, line width=.4mm] table[x expr=\thisrowno{0}, y expr=\thisrowno{131}*1000*1000*1000/\thisrowno{128}/15360/15360, header=true, col sep=comma] {data/clv_aligned_hotspots_barriers_memdp.csv};
    \addlegendentry{pdv01}
    
    \end{axis}

\end{tikzpicture}%

        \end{adjustbox}
    \end{subfigure}
    \caption{Code balance of the loops inside the hotspot functions of \clv\ on an Intel Ice Lake SP server. The dotted gray line indicates the end of the first ccNUMA domain. The data points represent the median out of ten separate runs with error bars omitted due to negligible fluctuations of 3.6\,\%. Note the different y-axis scales in both subfigures.}\label{fig:code_balance_naive}
\end{figure*}
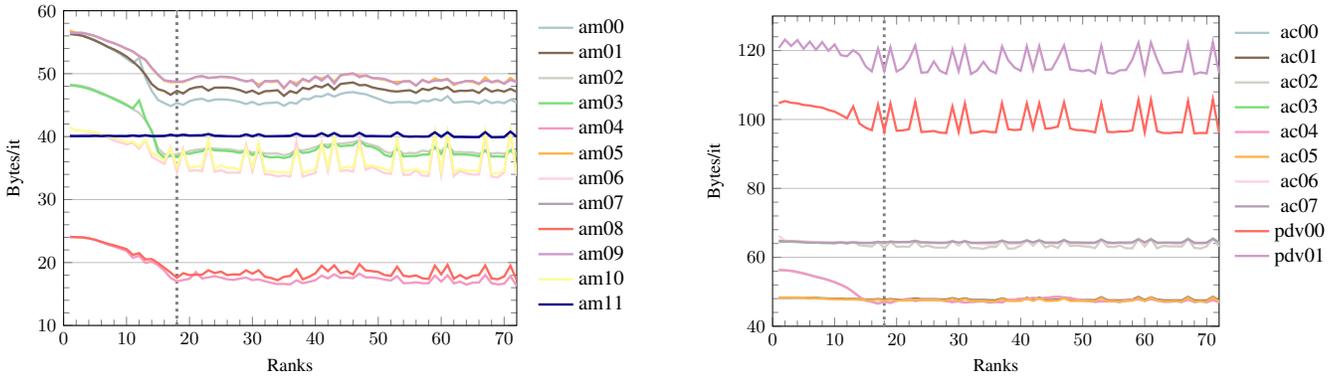

Among the 22 kernels, we can identify three classes of scaling behavior:
\begin{itemize}
    \item[(i)] Kernels with a strong reduction in code balance within the first ccNUMA domain and a constant code balance for the rest of the node and also a strong relative appearance of the prime number effect (am04, am06, am08, am10),
    \item[(ii)] Kernels with a slight reduction in code balance within the first ccNUMA domain, a constant code balance for the other ccNUMA domains, as well as a small appearance of the prime number effect (am00--am03, am05, am09, ac00, ac04, pdv00--01),
    \item[(iii)] Kernels with a constant code balance throughout all ranks of the node (am07, am11, ac03, ac07)
\end{itemize}
This matches the observation of the number of write-allocations required per kernel:
While for the loops of class (iii) there is no opportunity to reduce the data volume via non-temporal stores (because they do not have write misses), loops of class (i) could get rid of one WA element per iteration, and loops of class  (ii) could save two WAs per iteration in theory, i.e., if the data access patterns of the accessed arrays allow it:
On x86 architectures, non-temporal stores are always vectorized (starting with SSE) and require a SIMD-aligned store address, i.e., one that is a multiple of the SIMD width. 

\subsection{Possible causes for the prime number effect}
\revon
\clv\ spreads the prime factors in the number of ranks as evenly as possible across both dimensions for a square grid, starting with the (outer) y-dimension. However, \revoff at a prime process count, the square domain can only be cut along one \revon dimension\revoff, which is \revon chosen as\revoff\ the (inner) x-dimension. Hence, two reasons for the peculiar performance breakdowns at a prime number of processes come to mind: broken layer conditions or increased communication overhead.
\revon Both explanations turn out to be wrong and, in fact, a new WA evasion feature described in Sec~\ref{ssec:speci2m} is the reason for this, but for completeness we will falsify these two assumptions first. \revoff

If the outer dimension of the grid is cut when the number of ranks is prime, the LC with the full row size of the domain must be fulfilled for every process separately, tightening the LC as the number of processes increases~\cite{Stengel2015}. However, with the specific problem size used here, the required cache size for one array is easily available via the aggregate L2 and L3 cache sizes \emph{per core} (2.75\,\MiB). Furthermore, the code is written in a way that the \emph{inner} dimension is cut when the number of ranks is prime, hence the layer condition is the same for one rank and for any prime number of ranks. In conclusion, broken LCs cannot be responsible for increased memory traffic at prime process counts.

One-dimensional domain decomposition certainly has an adverse effect on communication overhead and might lead to increased memory traffic. To rule out that this is the reason for the performance drops, we create a breakdown of the program profile including MPI call times (see Fig.~\ref{fig:mpi}).
It shows the relative time spent either executing code or any of the MPI functions called.
While the MPI share of the code is in fact smaller when using only one ccNUMA domain, and the time spent in \texttt{MPI\_Allreduce} and \texttt{MPI\_Waitall} increases for a larger number of cores, this accounts only for a few percent of the overall runtime (note that the $y$ scale starts at 94\,\%) and also affects non-prime numbers with only few prime factors such as $38=2\times 19$.
We can therefore rule out both broken LCs and MPI communication to be the reason for the strange behavior visible at a prime number of cores.


Taking all observations into account, we must conclude that the prime number effect and the related drop of code balance across the first ccNUMA domain must be connected to write-allocate evasion and its occasional failure.
\begin{figure}[tp]
	\centering
	%
	\begin{tikzpicture}
    \centering
    \begin{axis}[
        height=5cm,
        width=\linewidth,
        ybar stacked,
        ymajorgrids=true,
        minor tick num=0,
        xlabel={Ranks},
        ylabel={runtime [\%]},
        ymin=94,
        ymax=100,
        xtick=data,
        scaled ticks=false,
        xticklabels = {2, 17, 18, 19, 37, 38, 71, 72},
        ytick = {94,95,96,97,98,99,100},
        legend pos=outer north east,
        legend style={draw=none, legend columns=3, at={(0.0, 1.35)}},
        label style={font=\small},
        tick label style={font=\small},
    ]
    
    \addplot[draw=blue-seq-4, fill=blue-seq-4, line width=.4mm] table[x expr=\coordindex, y expr=100*\thisrowno{2}/\thisrowno{1}, header=true, col sep=comma] {data/mpi.csv};
    \addlegendentry{Serial}
    \addplot[draw=red-seq-5, fill=red-seq-5, line width=.4mm] table[x expr=\coordindex, y expr=100*\thisrowno{4}/\thisrowno{1}, header=true, col sep=comma] {data/mpi.csv};
    \addlegendentry{MPI Waitall}
    \addplot[draw=red-seq-4, fill=red-seq-4, line width=.4mm] table[x expr=\coordindex, y expr=100*\thisrowno{4}/\thisrowno{1}, header=true, col sep=comma] {data/mpi.csv};
    \addlegendentry{MPI Allreduce}
    \addplot[draw=red-seq-3, fill=red-seq-3, line width=.4mm] table[x expr=\coordindex, y expr=100*\thisrowno{4}/\thisrowno{1}, header=true, col sep=comma] {data/mpi.csv};
    \addlegendentry{MPI Isend}
    \addplot[draw=red-seq-2, fill=red-seq-2, line width=.4mm] table[x expr=\coordindex, y expr=100*\thisrowno{4}/\thisrowno{1}, header=true, col sep=comma] {data/mpi.csv};
    \addlegendentry{MPI Reduce}
    \addplot[draw=red-seq-1, fill=red-seq-1, line width=.4mm] table[x expr=\coordindex, y expr=100*\thisrowno{4}/\thisrowno{1}, header=true, col sep=comma] {data/mpi.csv};
    \addlegendentry{MPI Barrier}
    \end{axis}

\end{tikzpicture}%

	\caption{Relative distribution of code execution and MPI time for different numbers of ranks.}
	\label{fig:mpi}
\end{figure}
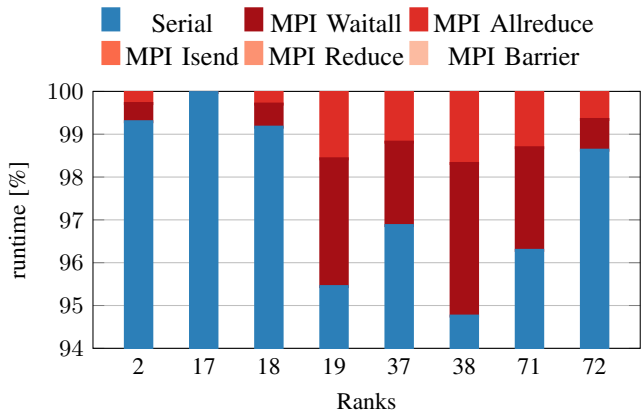


\section{Detailed analysis of write-allocate evasion} \label{sec:explanation}
\subsection{WA evasion via SpecI2M} \label{ssec:speci2m}
With the Ice Lake SP microarchitecture, Intel introduced a new data transaction type called ``SpecI2M''~\cite{papazian2020new}.
In contrast to the static store instructions with non-temporal hint, SpecI2M provides a dynamic-adaptive mechanism to reduce write-allocate traffic when memory bandwidth utilization is high. This means that this WA evasion mechanism is ineffective with serial code but kicks in when approaching bandwidth saturation within a memory domain. According to SpecI2M-related hardware performance events, cache lines eligible for SpecI2M are claimed at the L3 cache without a prior read-for-ownership (RFO). This sums up the publicly available information about SpecI2M. 

SpecI2M is controlled by an MSR bit that is disclosed by Intel under NDA. 
Running \clv\ with SpecI2M switched off, the code balance of all hotspot loops stays at the initial single-core value reported in the $\textrm{\BIT}_{\textrm{meas,1}}$ column of Table~\ref{tab:clv-analysis}, and the prime number effect is all but gone. 

The SpecI2M feature can be observed and validated when looking at isolated store microbenchmarks, shown in Figure~\ref{fig:stores}.
\begin{figure}[tp]
	\centering
	%
	\begin{tikzpicture}
    \centering
    \begin{axis}[
        height=6.5cm,
        width=0.8\linewidth,
        ymajorgrids=true,
        minor tick num=4,
        xlabel={Cores},
        ylabel={True traffic/reported traffic},
        ymin=0.9,
        ymax=2.1,        
        xmin=0,
        xmax=72,
        xtick = {0,10,20,30,40,50,60,70},
        legend pos=north east,
        legend style={draw=none, legend columns=1, at={(1.45, 1.0)}},
        label style={font=\small},
        tick label style={font=\small},
    ]
    \draw[dotted, draw=gray, line width=.5mm,] (18, 0) -- (18, 130);
    \draw[dotted, draw=gray, line width=.5mm,] (36, 0) -- (36, 130);
    \draw[dotted, draw=gray, line width=.5mm,] (54, 0) -- (54, 130);

    \addplot[draw=cbf-0, line width=.4mm, error bars/.cd, y dir=both, y explicit, error mark options={draw=cbf-0}] table[x expr=\thisrow{cores}, y expr=\thisrow{st_ratio_median}, y error plus expr=\thisrow{st_ratio_errp},y error minus expr=\thisrow{st_ratio_errm}, header=true, col sep=comma] {data/st1.csv};
    \addlegendentry{ST-1}
    \addplot[draw=cbf-1, line width=.4mm, error bars/.cd, y dir=both, y explicit, error mark options={draw=cbf-1}] table[x expr=\thisrow{cores}, y expr=\thisrow{st_ratio_median}, y error plus expr=\thisrow{st_ratio_errp},y error minus expr=\thisrow{st_ratio_errm}, header=true, col sep=comma] {data/st2.csv};
    \addlegendentry{ST-2}
    \addplot[draw=cbf-2, line width=.4mm, error bars/.cd, y dir=both, y explicit, error mark options={draw=cbf-2}] table[x expr=\thisrow{cores}, y expr=\thisrow{st_ratio_median}, y error plus expr=\thisrow{st_ratio_errp},y error minus expr=\thisrow{st_ratio_errm}, header=true, col sep=comma] {data/st3.csv};
    \addlegendentry{ST-3}

    \addplot[draw=cbf-5, line width=.4mm, error bars/.cd, y dir=both, y explicit, error mark options={draw=cbf-5}] table[x expr=\thisrow{cores}, y expr=\thisrow{st_ratio_median}, y error plus expr=\thisrow{st_ratio_errp},y error minus expr=\thisrow{st_ratio_errm}, header=true, col sep=comma] {data/stnt1.csv};
    \addlegendentry{ST-NT-1}
    \addplot[draw=cbf-6, line width=.4mm, error bars/.cd, y dir=both, y explicit, error mark options={draw=cbf-6}] table[x expr=\thisrow{cores}, y expr=\thisrow{st_ratio_median}, y error plus expr=\thisrow{st_ratio_errp},y error minus expr=\thisrow{st_ratio_errm}, header=true, col sep=comma] {data/stnt2.csv};
    \addlegendentry{ST-NT-2}
    \addplot[draw=cbf-7, line width=.4mm, error bars/.cd, y dir=both, y explicit, error mark options={draw=cbf-7}] table[x expr=\thisrow{cores}, y expr=\thisrow{st_ratio_median}, y error plus expr=\thisrow{st_ratio_errp},y error minus expr=\thisrow{st_ratio_errm}, header=true, col sep=comma] {data/stnt3.csv};
    \addlegendentry{ST-NT-3}


    
    \end{axis}

\end{tikzpicture}%

	\caption{Store ratio, i.e., ratio of actual memory traffic vs.\ explicitly initiated traffic with microbenchmarks using 1--3 store streams on an Intel Ice Lake SP server processor. ``NT'' marks benchmarks using non-temporal stores. The dotted gray lines indicate the end of each ccNUMA domain.}
	\label{fig:stores}
\end{figure}
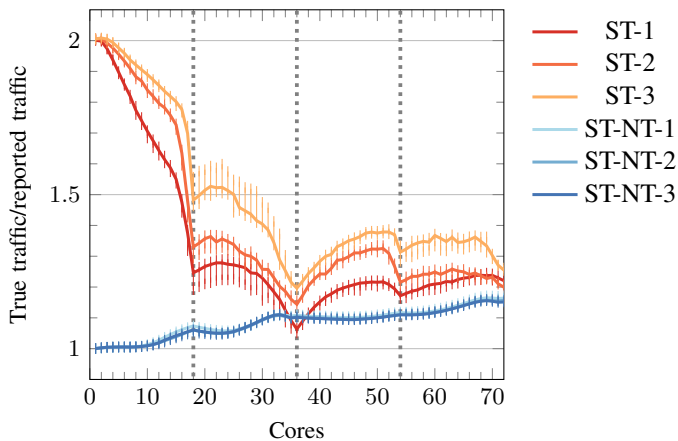
Here we store 10\,\GB\ of aligned array data as one, two, or three independent streams into main memory and use either normal stores or non-temporal~(NT) stores. The ratio of actual memory traffic to the pure store data volume (i.e., not including WAs) is reported.
While NT stores can avoid all WAs in the beginning, the store ratio increases from $1.0$ (i.e., all WA are avoided) to 1.16 (ST-NT-2, ST-NT-3) and 1.17 (ST-NT-1), respectively, when scaling to the full node; i.e., 16--17\% of the NT stores still trigger a WA\@.
\revon Moreover, we do not see any different behavior for these pure NT store benchmarks for SpecI2M activated or deactivated. \revoff\
The standard stores, on the other hand, independently of the number of streams, all have a store ratio of 2.0 with a single core, i.e., each store requires a WA\@.
Nearing saturation of the memory bandwidth on a ccNUMA domain, the ratio decreases steeply, which corroborates the assumption that effective WA evasion requires significant bandwidth draw. This is also why the store ratio rises again when a new ccNUMA domain is touched. 
One can also observe that the effectiveness of SpecI2M diminishes with a growing number of store streams.
While the best bandwidth is achieved at a full socket (36 cores) with a ratio of $1.06$, we see $1.2$--$1.25$ at a full node, i.e., 75--80\% of all WAs can be evaded by SpecI2M.

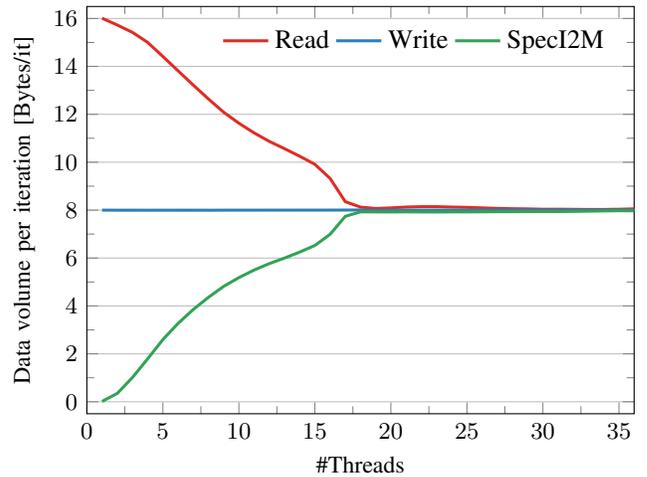
\begin{figure}[tp]
	\centering
	%
	\begin{tikzpicture}
    \centering
    \begin{axis}[
        height=7cm,
        width=\linewidth,
        ymajorgrids=true,
        minor tick num=1,
        xlabel={\#Threads},
        ylabel={Data volume per iteration [Bytes/it]},
        ymin=-0.5,
        ymax=16.5,
        xmin=0,
        xmax=36,
        ytick = {0, 2, 4, 6, 8, 10, 12, 14, 16},
        xtick = {0,5,10,15,20,25,30,35},
        legend pos=north east,
        legend style={draw=none, legend columns=3},
        label style={font=\small},
        tick label style={font=\small},
    ]

        \addplot[draw=red-seq-4, line width=.4mm] table[x expr=\thisrowno{0}, y expr=\thisrowno{1}*1000000000/\thisrowno{4}/\thisrowno{5}/\thisrowno{0}/16, header=true, col sep=comma] {data/copy_avx.csv};        
        \addlegendentry{Read}
        \addplot[draw=blue-seq-4, line width=.4mm] table[x expr=\thisrowno{0}, y expr=\thisrowno{2}*1000000000/\thisrowno{4}/\thisrowno{5}/\thisrowno{0}/16, header=true, col sep=comma] {data/copy_avx.csv};
        \addlegendentry{Write}
        \addplot[draw=green-seq-4, line width=.4mm] table[x expr=\thisrowno{0}, y expr=\thisrowno{3}*1000000000/\thisrowno{4}/\thisrowno{5}/\thisrowno{0}/16, header=true, col sep=comma] {data/copy_avx.csv};
        \addlegendentry{SpecI2M}
    
    \end{axis}

\end{tikzpicture}%

	\caption{Data volume for read and write operations as well as avoided WAs  related to SpecI2M of an array copy kernel for a single iteration updating one double-precision floating-point array entry.}
	\label{fig:copy_speci2m}
\end{figure}

Figure~\ref{fig:copy_speci2m} shows the increased effectiveness of SpecI2M with growing thread count using hardware performance counter measurements (LIKWID event for ICX and SPR: {\tt TOR\_INSERTS\_IA\_ITOM} for CBOXes/CHAs\footnote{Official event name from Intel for ICX and SPR: {\tt UNC\_CHA\_TOR\_INSERTS.IA\_ITOM}}) and a microbenchmark. The benchmark performs array copy operations (\verb.a(:) = b(:).) of two double-precision floating-point arrays with varying thread counts.
With a single thread, the write operations, each updating one array entry, require WAs, thus 16 \bytes\ are read from memory in one iteration. With 17 active hardware threads, the WAs are almost fully evaded by SpecI2M. Measurements on different Intel ``Ice Lake SP'' Xeon SKUs showed a different behavior.\footnote{\url{https://blogs.fau.de/hager/archives/8997}}

\subsection{SpecI2M in \clv}
With a phenomenological factor of $1.2$ for store streams applicable to WA evasion, one can build a refined model for the hotspot loops of the \clv\ mini-app. This is shown in Figure~\ref{fig:perf-model}.
\begin{figure*}[tp]
	\centering
	%
	\begin{tikzpicture}
    \centering    
    \begin{axis}[
        height=6.5cm,
        width=1.0\linewidth,
        ybar,
        bar width=0.2cm,
        ymajorgrids=true,
        xlabel={Loop},
        ylabel={Code Balance [\BIT]},
        ymin=0,
        ymax=116,
        xmin=-0.5,
        xmax=21.5,
        xtick=data,
        scaled ticks=false,
        xticklabels = {am00,am01,am02,am03,am04,am05,am06,am07,am08,am09,am10,am11,ac00,ac01,ac02,ac03,ac04,ac05,ac06,ac07,pdv00,pdv01},
        x tick label style={rotate=45,anchor=east},
        ytick = {0, 8, 16, 24, 32, 40, 48, 56, 64, 72, 80, 88, 96, 104, 112, 120},
        reverse legend=true,
        legend pos=north west,
        legend style={draw=none, legend columns=2,},
        label style={font=\small},
        tick label style={font=\small},
    ]
    
    \addplot[draw=blue-seq-4, fill=blue-seq-4, line width=.4mm] table[x expr=\coordindex, y expr=\thisrowno{3}, header=true] {data/model.csv};
    \addlegendentry{Measured} \label{plot:perf}
    \addplot[draw=blue-seq-3, fill=blue-seq-3, line width=.4mm] table[x expr=\coordindex, y expr=\thisrowno{5}, header=true] {data/model.csv};
    \addlegendentry{Optimized} \label{plot:perf-opt}
    
    \end{axis}

    \begin{axis}[
        height=6.5cm,
        width=1.0\linewidth,
        ymajorgrids=false,
        xlabel={},
        ylabel={},
        ymin=0,
        ymax=116,
        xmin=-0.5,
        xmax=21.5,
        scaled ticks=false,
        ticks=none,
        xtick={},
        ytick={},
        reverse legend=true,
        legend pos=north west,
        legend style={draw=none, legend columns=2,},
        label style={font=\small},
        tick label style={font=\small},
    ]
    \addplot[draw=red-seq-3, line width=.5mm,] coordinates {(-0.4, 24+16*1.20) (0.4, 
    24+16*1.20)}; 
    \addlegendentry{Prediction} \label{plot:pred}
    \draw[draw=red-seq-3, line width=.5mm,] (0.6, 24+16*1.20) -- (1.4, 24+16*1.20); 
    \draw[draw=red-seq-3, line width=.5mm,] (1.6, 16+16*1.20) -- (2.4, 16+16*1.20); 
    \draw[draw=red-seq-3, line width=.5mm,] (2.6, 16+16*1.20) -- (3.4, 16+16*1.20); 
    \draw[draw=red-seq-3, line width=.5mm,] (3.6, 16+1.6) -- (4.4, 16+1.6); 
    \draw[draw=red-seq-3, line width=.5mm,] (4.6, 24+16*1.20) -- (5.4, 24+16*1.20); 
    \draw[draw=red-seq-3, line width=.5mm,] (5.6, 32+1.6) -- (6.4, 32+1.6); 
    \draw[draw=red-seq-3, line width=.5mm,] (6.6, 40) -- (7.4, 40); 
    \draw[draw=red-seq-3, line width=.5mm,] (7.6, 16+1.6) -- (8.4, 16+1.6); 
    \draw[draw=red-seq-3, line width=.5mm,] (8.6, 24+16*1.20) -- (9.4, 24+16*1.20); 
    \draw[draw=red-seq-3, line width=.5mm,] (9.6, 32+1.6) -- (10.4, 32+1.6); 
    \draw[draw=red-seq-3, line width=.5mm,] (10.6, 40) -- (11.4, 40); 
    \draw[draw=red-seq-3, line width=.5mm,] (11.6, 24+16*1.20) -- (12.4, 24+16*1.20); 
    \draw[draw=red-seq-3, line width=.5mm,] (12.6, 16+16*1.20) -- (13.4, 16+16*1.20); 
    \draw[draw=red-seq-3, line width=.5mm,] (13.6, 32+16*1.20) -- (14.4, 32+16*1.20); 
    \draw[draw=red-seq-3, line width=.5mm,] (14.6, 64) -- (15.4, 64); 
    \draw[draw=red-seq-3, line width=.5mm,] (15.6, 24+16*1.20) -- (16.4, 24+16*1.20); 
    \draw[draw=red-seq-3, line width=.5mm,] (16.6, 16+16*1.20) -- (17.4, 16+16*1.20); 
    \draw[draw=red-seq-3, line width=.5mm,] (17.6, 32+16*1.20) -- (18.4, 32+16*1.20); 
    \draw[draw=red-seq-3, line width=.5mm,] (18.6, 64) -- (19.4, 64); 
    \draw[draw=red-seq-3, line width=.5mm,] (19.6, 72+16*1.20) -- (20.4, 72+16*1.20); 
    \draw[draw=red-seq-3, line width=.5mm,] (20.6, 88+16*1.20) -- (21.4, 88+16*1.20); 

    \addplot[draw=red-seq-5, line width=.5mm,] coordinates {(-0.4, 40) (0.5, 40)}; 
    \addlegendentry{Prediction min.} \label{plot:pred-min}
    \draw[draw=red-seq-5, line width=.5mm,] (0.6, 40) -- (1.4, 40); 
    \draw[draw=red-seq-5, line width=.5mm,] (1.6, 32) -- (2.4, 32); 
    \draw[draw=red-seq-5, line width=.5mm,] (2.6, 32) -- (3.4, 32); 
    \draw[draw=red-seq-5, line width=.5mm,] (3.6, 16) -- (4.4, 16); 
    \draw[draw=red-seq-5, line width=.5mm,] (4.6, 40) -- (5.4, 40); 
    \draw[dotted,draw=red-seq-5, line width=.5mm,] (5.6, 32) -- (6.4, 32); 
    \draw[dotted,draw=red-seq-5, line width=.5mm,] (6.6, 40) -- (7.4, 40); 
    \draw[draw=red-seq-5, line width=.5mm,] (7.6, 16) -- (8.4, 16); 
    \draw[draw=red-seq-5, line width=.5mm,] (8.6, 40) -- (9.4, 40); 
    \draw[draw=red-seq-5, line width=.5mm,] (9.6, 32) -- (10.4, 32); 
    \draw[dotted,draw=red-seq-5, line width=.5mm,] (10.6, 40) -- (11.4, 40); 
    \draw[draw=red-seq-5, line width=.5mm,] (11.6, 40) -- (12.4, 40); 
    \draw[draw=red-seq-5, line width=.5mm,] (12.6, 32) -- (13.4, 32); 
    \draw[draw=red-seq-5, line width=.5mm,] (13.6, 48) -- (14.4, 48); 
    \draw[dotted,draw=red-seq-5, line width=.5mm,] (14.6, 64) -- (15.4, 64); 
    \draw[draw=red-seq-5, line width=.5mm,] (15.6, 40) -- (16.4, 40); 
    \draw[draw=red-seq-5, line width=.5mm,] (16.6, 32) -- (17.4, 32); 
    \draw[draw=red-seq-5, line width=.5mm,] (17.6, 48) -- (18.4, 48); 
    \draw[dotted,draw=red-seq-5, line width=.5mm,] (18.6, 64) -- (19.4, 64); 
    \draw[draw=red-seq-5, line width=.5mm,] (19.6, 88) -- (20.4, 88); 
    \draw[draw=red-seq-5, line width=.5mm,] (20.6, 104) -- (21.4, 104); 

    \addlegendimage{/pgfplots/refstyle=plot:perf}\addlegendentry{Original measurement}
    \addlegendimage{/pgfplots/refstyle=plot:perf-opt}\addlegendentry{Optimized measurement}
    
    \end{axis}

\end{tikzpicture}%

	\caption{Predicted minimal code balance and the refined performance model (including a phenomenological SpecI2M factor for writes) compared to the full-node measured code balance per loop of the original code and an optimized version using non-temporal stores and manual rearrangement of the code in loops ac01 and ac05. The latter version achieves on average 5.8\% lower code balance with a maximum of 23.2\%.}
	\label{fig:perf-model}
\end{figure*}
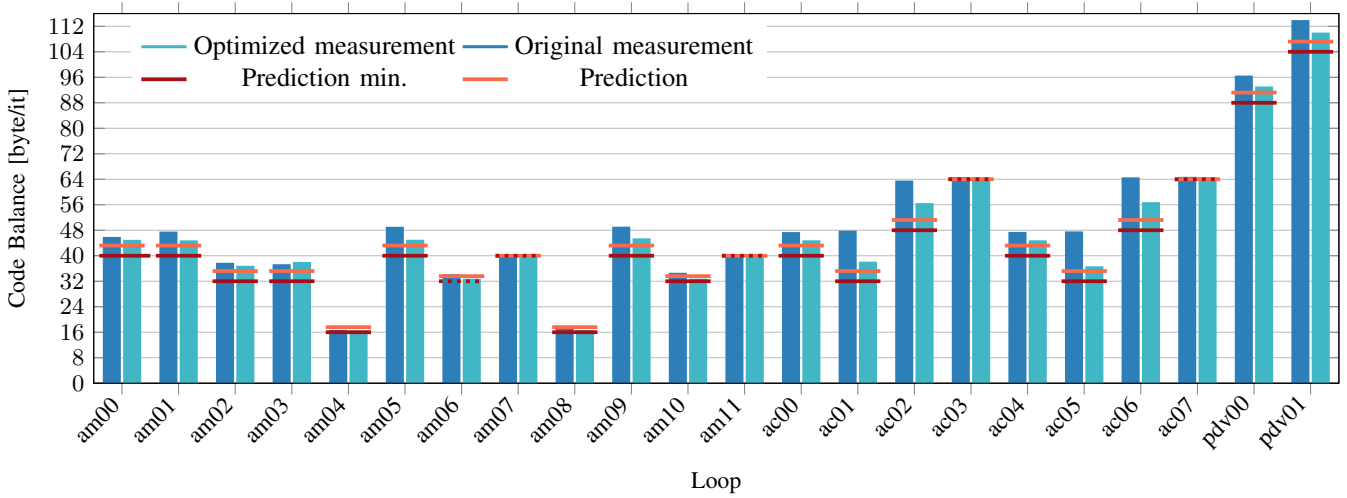
While ``Original Measurement'' refers to the original measurement with 72~cores (i.e., the data point for 72~cores in Figure~\ref{fig:code_balance_naive} for each kernel), ``Prediction min.'' is the minimum code balance with no WA.
``Prediction'' shows the refined model with the phenomenological SpecI2M factor of $1.2$ from Figure~\ref{fig:stores} applied to all loops with one or more write streams with possible WA avoidance.
Note that this factor is machine dependent and might differ for a different Ice Lake SP server model.
With an average error of 7\,\% of the actual code balance, the model is already very close to the measurement.
However, for loops am04 and am08 the code balance prediction is slightly higher than the measurement.
It seems here the SpecI2M feature works particularly well since both loops show the same structure in data accesses throughout the grid (loading four neighbor cells from the input grid while storing in a single output cell).
Furthermore, for ac01, ac02, ac05, and ac06 the model only predicts 74--81\% of the measurement, which actually matches a prediction without SpecI2M.
While loops ac02 and ac06 are more complex and contain conditional branching, thus, the hardware might not be able to identify the stores to be eligible for SpecI2M, the other two loops are very simple and contain only a memory copy and a store of the previously loaded value, which happens in almost the same way in other loops (am02, am03).
We found that a simple reorganization of these loops which does not change the semantics but actually creates a (recoverable) read-after-write dependency, made the hardware apply SpecI2M.
We were not be able to pinpoint the exact reason why in these cases SpecI2M does not work as expected. We could rule out several hypotheses like unaligned arrays, the use of masking registers within the peeling loop, or a different usage of assembly instructions.

As shown in Figure~\ref{fig:stores}, the explicit use of NT stores is often more efficient for avoiding WAs. To get the best possible performance, we therefore added compiler directives (\texttt{!DIR\$ vector nontemporal}) at each hotspot loop and applied the described restructuring of loops ac01 and ac05 to allow the use of SpecI2M.
While the NT store directive makes the compiler use non-temporal stores for only one of the write streams (if there are multiple) because of the address alignment constraint, the other stream can be optimized by SpecI2M. As a result, we observe an average improvement compared to the original version of 5.8\%, shown as ``Optimized Measurement'' in Fig.\ref{fig:perf-model}
For this version, our code balance prediction has an error of 1\%.

\subsection{Prime number effect}

With accurate traffic predictions for single-core and full-node executions in place, we still need to explain the significant rise in code balance when using a prime number of processes.
This prime number effect is due to two different phenomena which are both connected to the short inner loop when iterating over the grid.
With a prime number of ranks, only the inner dimension of the grid is cut, leading local inner domain sizes between 216 (in case of 71 processes) and 809 (in case of 19 processes) elements after the first ccNUMA domain. Non-prime runs have at least 1920 and up to 7680 elements per row (excluding the case of a single rank with obviously 15360 elements per row).
The smaller the inner dimension, the more extra traffic is (relatively) caused by the halo layers in front and after each row (five elements in total). This increase is present independent of SpecI2M being active or not.
For an inner dimension of 216, one additional cache line, i.e., eight double-precision elements, results in $8/(216+8)=3.57\%$ of extra data volume per read stream.
In fact, when looking at the increase of read data volume for prime numbers after the first ccNUMA domain of all hotspot loops with SpecI2M deactivated, we can measure an average increase between 1.33\% and 3.54\% per loop, matching the prediction perfectly.
Since the elements at the start and the end of a row are not always aligned to the start and the end of a cache line, the processor must WA these partial cache lines and thus creates the same additional data volume as for the reads of up to one cache line per row.
Again we can see this when measuring the write data volume with deactivated SpecI2M and observe an increase of write data volume of 1.09\% on average, with a maximum of 2.66\%.

Despite this general phenomenon of increased data volume for prime numbers, we still see an additional overhead of up to 24\% of the read data volume for loops of class (i), i.e., am04, am06, am08, and am10, and an average of additionally 6\% of read data volume with up to 10\% for the pdv loops.
Although the write data volume is unaffected, this massive growth of read data volume seems to only affect the loops with one write stream amenable to WA evasion and is related to the SpecI2M feature in combination with short loops and incomplete cache line writes. One feature that can further increase the read data volume is the ``adjacent cache line prefetch,'' which transfers an additional cache line from memory upon any cache miss, effectively doubling the cache line size. However, the corresponding extra traffic is by far not sufficient to deliver a quantitative explanation of the prime number effect.

One may speculate that incomplete cache line writes trigger prefetch operations into the L3 cache, with a prefetch distance large enough to explain the extra traffic. We conduct an experiment with a simple microbenchmark to test this hypothesis with the results shown in Figure~\ref{fig:copy}.
\begin{figure}[tp]
	\centering
	%
	\begin{tikzpicture}
    \centering
    \begin{axis}[
        height=7cm,
        width=\linewidth,
        ymajorgrids=true,
        minor tick num=1,
        xlabel={Halo size},
        ylabel={Memory read/write ratio},
        ymin=1,
        ymax=2,
        xmin=-1,
        xmax=18,
        ytick = {1, 1.2, 1.4, 1.6, 1.8, 2},
        xtick = {0, 2, 4, 6, 8, 10, 12, 14, 16},
        legend pos=outer north east,
        legend style={draw=none, legend columns=3, at={(0.0, 1.25)}},
        label style={font=\small},
        tick label style={font=\small},
    ]

        \addplot[draw=blue-seq-5, line width=.4mm] table[x expr=\thisrowno{1}, y expr=\thisrowno{7}/\thisrowno{9}, header=true, col sep=comma] {data/copy216.csv};
        \addlegendentry{216}
        \addplot[draw=red-seq-5, line width=.4mm] table[x expr=\thisrowno{1}, y expr=\thisrowno{7}/\thisrowno{9}, header=true, col sep=comma] {data/copy530.csv};
        \addlegendentry{530}
        \addplot[draw=green-seq-5, line width=.4mm] table[x expr=\thisrowno{1}, y expr=\thisrowno{7}/\thisrowno{9}, header=true, col sep=comma] {data/copy1920.csv};
        \addlegendentry{1920}
        
        %

        \addplot[draw=blue-seq-3, line width=.4mm] table[x expr=\thisrowno{1}, y expr=\thisrowno{7}/\thisrowno{9}, header=true, col sep=comma] {data/copy216-pfoff.csv};
        \addlegendentry{216 PF off}
        \addplot[draw=red-seq-3, line width=.4mm] table[x expr=\thisrowno{1}, y expr=\thisrowno{7}/\thisrowno{9}, header=true, col sep=comma] {data/copy530-pfoff.csv};
        \addlegendentry{530 PF off}
        \addplot[draw=green-seq-2, line width=.4mm] table[x expr=\thisrowno{1}, y expr=\thisrowno{7}/\thisrowno{9}, header=true, col sep=comma] {data/copy1920-pfoff.csv};
        \addlegendentry{1920 PF off}
    
    \end{axis}

\end{tikzpicture}%

	\caption{Read-to-write data volume ratio of a copy microbenchmark (\texttt{a(:) = b(:)}) over different halo sizes on a full Ice Lake SP server node. Each benchmark copies 76.8\,\GB\ of data in batches of either 216, 530, or 1920 elements. ``PF off'' indicates that all hardware prefetchers were turned off. Error bars are omitted due to deviations less than 5\%.}\label{fig:copy}
\end{figure}
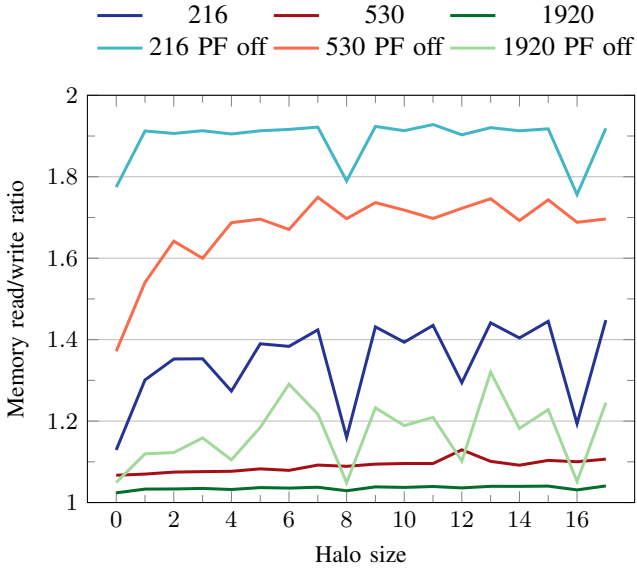
We simulated the effectiveness of SpecI2M with different inner loop lengths and halo sizes; 216 corresponds to the inner dimension when the grid is distributed across 71 ranks, 530 corresponds to 29 ranks, and 1920 to the inner dimension for 64 and 72 ranks, respectively, which is the smallest size of the inner dimensions for non-prime numbers, while all other prime numbers larger than 7 lead to smaller dimensions.
The data shows the read-to-write ratio of a simple copy microbenchmark (\verb.a(:) = b(:).) with different halo sizes; starting with 0, meaning no halo layer, up to 17 for more than two cache lines of halo.
A ratio of 1.0 would indicate that all WAs from the destination array can be avoided, while a ratio of 2.0 means that every element of the target has to be read first.

For the sizes 216 and 1920, which are multiples of the cache line size, a significant amount of WAs can be avoided if the code can write full cache lines in every or every other iteration (with a halo size of 4, 8, 12, or 16), while this does not show up clearly for 530, corresponding to 66.25 cache lines.
This effect can still be observed when turning off all hardware prefetchers (``PF off'').
However, the more important effect in terms of extra data volume comes from the length of the inner dimension.
While copying the array in batches of 216 elements only leads to an average  ratio of 1.35, a batch size of of 530 shows an average ratio of 1.09, and lines of 1920 elements decrease this ratio even further to 1.04.
Furthermore, one can observe that this trend withstands the absence of any prefetchers, even though the read-to-write ratio increases drastically, especially for a halo size leading to partial cache lines to be written.
This means that, contrary to our first hypothesis, active prefetchers and long loops (hundreds of cache lines) benefit the effect of SpecI2M, while short loops are detrimental to its effectiveness.

We can thus conclude that even though prefetchers influence the SpecI2M behavior, they do so in a beneficial way for generic streaming codes.
However, the more significant impact on the performance is created by the length of the inner loop, with SpecI2M working drastically worse on short loops and with an increase of efficiency when enlarging the inner dimension.
SpecI2M might get disturbed by various effects such as the conditional jumps, different instructions for the loop control or eventual peeling and remainder loops.

\subsection{SpecI2M on Sapphire Rapids (SPR)} \label{ssec:spr}
To evaluate the efficiency of SpecI2M on Intel's most modern CPU architecture, we redo the\revon\ previous\revoff\ experiments on \revon\ Sapphire Rapids servers.
\revon\ For an initial insight on the impact of Sub-NUMA Clustering (SNC) on SpecI2M, we reproduce the store ratio benchmark (see Fig.~\ref{fig:stores}) on a 104-core Intel Platinum 8470 server for SNC on and off as shown in Figure~\ref{fig:stores-spr-fritz}.

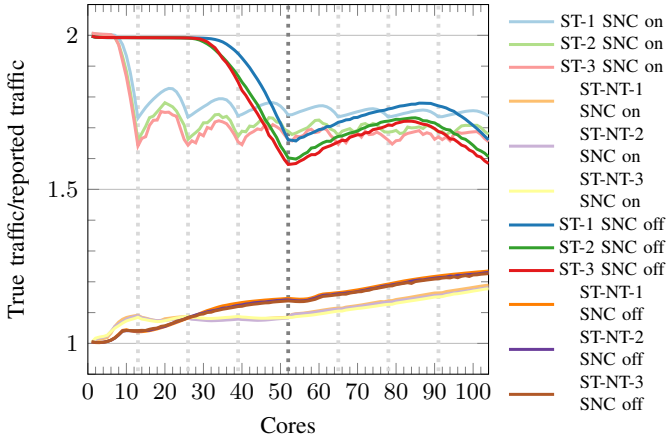
\begin{figure}[tp]
	\centering
	%
	\begin{tikzpicture}
    \centering
    \begin{axis}[
        height=6.5cm,
        width=0.78\linewidth,
        ymajorgrids=true,
        minor tick num=4,
        xlabel={Cores},
        ylabel={True traffic/reported traffic},
        ymin=0.9,
        ymax=2.1,        
        xmin=0,
        xmax=104,
        xtick = {0,10,20,30,40,50,60,70,80,90,100,110},
        legend pos=north east,
        legend style={nodes={scale=0.7, transform shape}, cells={align=left}, draw=none, legend columns=1, at={(1.47, 1.0)}},
        label style={font=\small},
        tick label style={font=\small},
    ]
    \draw[dotted, draw=light-gray, line width=.5mm,] (13, 0) -- (13, 130);
    \draw[dotted, draw=light-gray, line width=.5mm,] (26, 0) -- (26, 130);
    \draw[dotted, draw=light-gray, line width=.5mm,] (39, 0) -- (39, 130);
    \draw[dotted, draw=gray, line width=.5mm,] (52, 0) -- (52, 130);
    \draw[dotted, draw=light-gray, line width=.5mm,] (65, 0) -- (65, 130);
    \draw[dotted, draw=light-gray, line width=.5mm,] (78, 0) -- (78, 130);
    \draw[dotted, draw=light-gray, line width=.5mm,] (91, 0) -- (91, 130);

    \addplot[draw=longqual-0, line width=.4mm] table[x expr=\thisrowno{0}, y expr=\thisrowno{4}, header=true, col sep=comma] {data/rebuttal/st1-spr-fritz.csv};
    \addlegendentry{ST-1 SNC on}
    \addplot[draw=longqual-2, line width=.4mm] table[x expr=\thisrowno{0}, y expr=\thisrowno{4}, header=true, col sep=comma] {data/rebuttal/st2-spr-fritz.csv};
    \addlegendentry{ST-2 SNC on}
    \addplot[draw=longqual-4, line width=.4mm] table[x expr=\thisrowno{0}, y expr=\thisrowno{4}, header=true, col sep=comma] {data/rebuttal/st3-spr-fritz.csv};
    \addlegendentry{ST-3 SNC on}

    \addplot[draw=longqual-6, line width=.4mm] table[x expr=\thisrowno{0}, y expr=\thisrowno{4}, header=true, col sep=comma] {data/rebuttal/stnt1-spr-fritz.csv};
    \addlegendentry{ST-NT-1\\SNC on}
    \addplot[draw=longqual-8, line width=.4mm] table[x expr=\thisrowno{0}, y expr=\thisrowno{4}, header=true, col sep=comma] {data/rebuttal/stnt2-spr-fritz.csv};
    \addlegendentry{ST-NT-2\\SNC on}
    \addplot[draw=longqual-10, line width=.4mm] table[x expr=\thisrowno{0}, y expr=\thisrowno{4}, header=true, col sep=comma] {data/rebuttal/stnt3-spr-fritz.csv};
    \addlegendentry{ST-NT-3\\SNC on}

    \addplot[draw=longqual-1, line width=.4mm] table[x expr=\thisrowno{0}, y expr=\thisrowno{4}, header=true, col sep=comma] {data/rebuttal/st1-nps1-spr-fritz.csv};
    \addlegendentry{ST-1 SNC off}
    \addplot[draw=longqual-3, line width=.4mm] table[x expr=\thisrowno{0}, y expr=\thisrowno{4}, header=true, col sep=comma] {data/rebuttal/st2-nps1-spr-fritz.csv};
    \addlegendentry{ST-2 SNC off}
    \addplot[draw=longqual-5, line width=.4mm] table[x expr=\thisrowno{0}, y expr=\thisrowno{4}, header=true, col sep=comma] {data/rebuttal/st3-nps1-spr-fritz.csv};
    \addlegendentry{ST-3 SNC off}

    \addplot[draw=longqual-7, line width=.4mm] table[x expr=\thisrowno{0}, y expr=\thisrowno{4}, header=true, col sep=comma] {data/rebuttal/stnt1-nps1-spr-fritz.csv};
    \addlegendentry{ST-NT-1\\SNC off}
    \addplot[draw=longqual-9, line width=.4mm] table[x expr=\thisrowno{0}, y expr=\thisrowno{4}, header=true, col sep=comma] {data/rebuttal/stnt2-nps1-spr-fritz.csv};
    \addlegendentry{ST-NT-2\\SNC off}
    \addplot[draw=longqual-11, line width=.4mm] table[x expr=\thisrowno{0}, y expr=\thisrowno{4}, header=true, col sep=comma] {data/rebuttal/stnt3-nps1-spr-fritz.csv};
    \addlegendentry{ST-NT-3\\SNC off}

    \end{axis}

\end{tikzpicture}%

	\caption{\revon\ Store ratio, i.e., ratio of actual memory traffic vs.\ explicitly initiated traffic with microbenchmarks using 1--3 store streams on Intel Xeon Platinum 8470 Sapphire Rapids. ``NT'' marks benchmarks using non-temporal stores. ``SNC'' describes if the system was set in SNC mode. The dotted gray lines indicate the end of the ccNUMA domain (note that for SNC=off there is only one ccNUMA domain per socket). Error bars are omitted due to run-to-run fluctuations being less than 5\%.\revoff}
	\label{fig:stores-spr-fritz}
\end{figure}

While SpecI2M kicks in much faster in SNC=on mode due to the faster saturation of the (significantly smaller) ccNUMA domain, we observe that the overall efficiency on a full socket with SNC=off is better by 5\% compared to the same number of streams with SNC=on, respectively, with the best ratio being 1.58 on a full socket.
While the difference for the store benchmarks with NT-store hints is smaller than for standard stores, it is important to note that here the ccNUMA domain boundaries are not visible and that the NT stores are slightly more efficient with SNC activated.
However, since we care the most about SpecI2M in this work, we will focus on a system setting with Sub-NUMA Clustering off for the rest of this analysis.
\revoff

The results\revon\ for the store ratio and read-to-write data volume on the Intel Platinum 8480+ server\revoff\ can be found in Figure~\ref{fig:stores-spr} and Figure~\ref{fig:copy-spr}, respectively.
While the overall trend for the store microbenchmarks stays the same with SpecI2M kicking in after a while and worsening when starting a new ccNUMA domain, we can observe several differences:
In comparison with ICX, there is no decline in SpecI2M performance with growing number of store streams anymore but SpecI2M only shows any benefit after 18 cores used, while on Ice Lake there was an immediate effect after 3 cores.
Furthermore, SpecI2M only avoids roughly 50\% of the write-allocates in the best case of full sockets, while on Ice Lake the best ratio between actual and reported traffic attains 1.06 on one socket and 1.2--1.25 on  the full node.
\revon\ Moreover, comparing these results to the benchmark conducted on the Xeon Platinum 8470 node in Figure~\ref{fig:stores-spr-fritz}, where we see an even worse store ratio especially for a single store stream (66\% vs.\ 51\% of all WAs evaded), we must conclude that the SpecI2M efficiency varies between different server models of the same generation.\revoff
The ratio of data volume for non-temporal stores behaves in the same way as on the older CPU architecture and slowly increases with a growing number of cores to 1.18 (1.16--1.17 on Ice Lake).
This results in the assumption that---even though Intel improved the efficiency for multiple stores---the SpecI2M feature for store-heavy codes is less beneficial on Sapphire Rapids than on Ice Lake systems.

\begin{figure}[tp]
	\centering
	%
	\begin{tikzpicture}
    \centering
    \begin{axis}[
        height=6.5cm,
        width=0.78\linewidth,
        ymajorgrids=true,
        minor tick num=4,
        xlabel={Cores},
        ylabel={True traffic/reported traffic},
        ymin=0.9,
        ymax=2.1,        
        xmin=0,
        xmax=112,
        xtick = {0,10,20,30,40,50,60,70,80,90,100,110},
        legend pos=north east,
        legend style={draw=none, legend columns=1, at={(1.48, 1.0)}},
        label style={font=\small},
        tick label style={font=\small},
    ]
    \draw[dotted, draw=gray, line width=.5mm,] (56, 0) -- (56, 130);

    \addplot[draw=cbf-0, line width=.4mm] table[x expr=\thisrowno{0}, y expr=\thisrowno{4}, header=true, col sep=comma] {data/st1.spr.csv};
    \addlegendentry{ST-1}
    \addplot[draw=cbf-1, line width=.4mm] table[x expr=\thisrowno{0}, y expr=\thisrowno{4}, header=true, col sep=comma] {data/st2.spr.csv};
    \addlegendentry{ST-2}
    \addplot[draw=cbf-2, line width=.4mm] table[x expr=\thisrowno{0}, y expr=\thisrowno{4}, header=true, col sep=comma] {data/st3.spr.csv};
    \addlegendentry{ST-3}

    \addplot[draw=cbf-5, line width=.4mm] table[x expr=\thisrowno{0}, y expr=\thisrowno{4}, header=true, col sep=comma] {data/stnt1.spr.csv};
    \addlegendentry{ST-NT-1}
    \addplot[draw=cbf-6, line width=.4mm] table[x expr=\thisrowno{0}, y expr=\thisrowno{4}, header=true, col sep=comma] {data/stnt2.spr.csv};
    \addlegendentry{ST-NT-2}
    \addplot[draw=cbf-7, line width=.4mm] table[x expr=\thisrowno{0}, y expr=\thisrowno{4}, header=true, col sep=comma] {data/stnt3.spr.csv};
    \addlegendentry{ST-NT-3}
    
    \end{axis}

\end{tikzpicture}%

	\caption{Store ratio, i.e., ratio of actual memory traffic vs.\ explicitly initiated traffic with microbenchmarks using 1--3 store streams on\revon\ Intel Xeon Platinum 8480+\revoff\ Sapphire Rapids. ``NT'' marks benchmarks using non-temporal stores. The dotted gray line indicates the end of the ccNUMA domain. Error bars are omitted due to deviations less than 5\%.}
	\label{fig:stores-spr}
\end{figure}
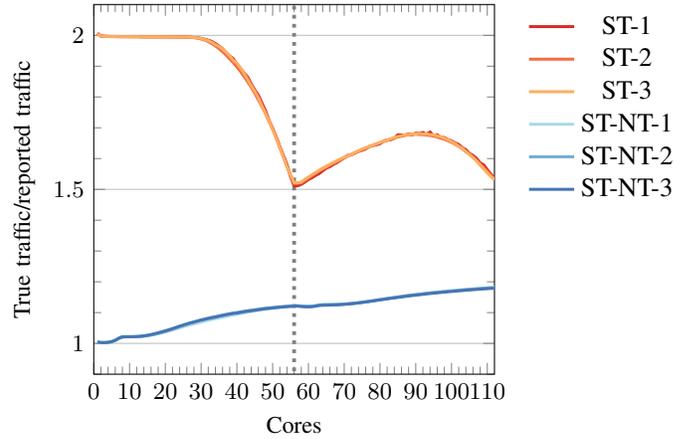

However, when looking at the copy microbenchmark with different halo sizes, we can observe that SpecI2M works better for the shortest dimensions of 216 elements. 
The read-to-write ratio slightly decreases by 3\% on average but with a significant difference of 10\% for the cases of no or an aligned halo layer (i.e., a halo size of 0, 8, or 16).
This indicates that---at least for SPR---the strip-mining of the loop (even with gaps of muiltiples of 8) does not impact the SpecI2M performance anymore, and inefficiencies of it rather result from accessing non-sequential or incomplete cache lines.
For dimensions of 530 and 1920, no difference larger than 1\% is visible between the two server generations.
In general, we can only observe an improvement of SpecI2M on Sapphire Rapids for very specific cases, while otherwise there is no evident change in the functionality of the feature or even a deterioration.

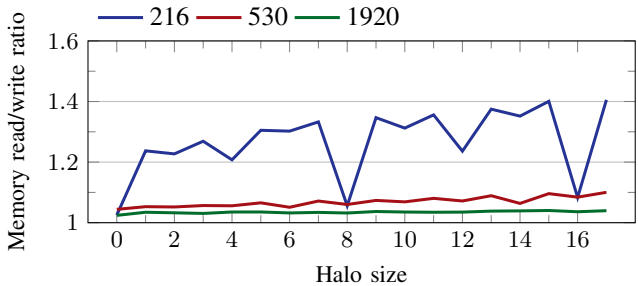
\begin{figure}[tp]
	\centering
	%
	\begin{tikzpicture}
    \centering
    \begin{axis}[
        height=4cm,
        width=\linewidth,
        ymajorgrids=true,
        minor tick num=1,
        xlabel={Halo size},
        ylabel={Memory read/write ratio},
        ymin=1,
        ymax=1.6,
        xmin=-1,
        xmax=18,
        ytick = {1, 1.2, 1.4, 1.6},
        xtick = {0, 2, 4, 6, 8, 10, 12, 14, 16},
        legend pos=outer north east,
        legend style={draw=none, legend columns=3, at={(0.0, 1.25)}},
        label style={font=\small},
        tick label style={font=\small},
    ]

        \addplot[draw=blue-seq-5, line width=.4mm] table[x expr=\thisrowno{1}, y expr=\thisrowno{7}/\thisrowno{9}, header=true, col sep=comma] {data/copy216.spr.csv};
        \addlegendentry{216}
        \addplot[draw=red-seq-5, line width=.4mm] table[x expr=\thisrowno{1}, y expr=\thisrowno{7}/\thisrowno{9}, header=true, col sep=comma] {data/copy530.spr.csv};
        \addlegendentry{530}
        \addplot[draw=green-seq-5, line width=.4mm] table[x expr=\thisrowno{1}, y expr=\thisrowno{7}/\thisrowno{9}, header=true, col sep=comma] {data/copy1920.spr.csv};
        \addlegendentry{1920}

    
    \end{axis}

\end{tikzpicture}%

	\caption{Read-to-write data volume ratio of a copy microbenchmark (\texttt{a(:) = b(:)}) over different halo sizes on a full Sapphire Rapids\revon\ 8480+\revoff\ server node. Each benchmark copies 76.8\,\GB\ of data in batches of either 216, 530, or 1920 elements.
    Error bars are omitted due to deviations less than 5\%.}\label{fig:copy-spr}
\end{figure}

\section{Conclusion and Future Work} \label{sec:conclusion}
Benchmarking the MPI version of the \clv\ code in the SPEChpc 2021 benchmark suite using the ``Tiny'' data set on an Intel Ice Lake SP (ICX) server, we discovered breakdowns in performance when the number of processes is prime, i.e., when the inner dimension of the local domain is short. These breakdowns were accompanied by spikes in code balance, specifically increased read traffic from main memory. After ruling out MPI overhead and broken layer conditions as reasons for this effect, we built detailed traffic models for the hotspot loops in \clv\ and showed that the code balance for single-core and full-node executions can be accurately modeled when taking the SpecI2M automatic write-allocate (WA) evasion feature of the ICX chip into account. We also showed that SpecI2M is generally less effective in WA evasion than non-temporal store instructions; the combination of both together with minimal code changes resulted in best performance and lowest code balance on the full node. The prime number effect, however, could be attributed to a failure of SpecI2M to avoid write-allocates when the inner domain size is short. While this connection is undisputed, we could not accurately model the effect quantitatively.
Analyzing SpecI2M on an Intel Sapphire Rapids (SPR) server, we could not find any major improvements compared to its predecessor and could even observe deterioration (i.e., less effective WA evasion), especially when using only a limited number of cores per ccNUMA domain. 

\revon
The findings of this paper on SpecI2M can not only be applied to the \clv\ application, but to any memory-bound code that writes data into main memory in a streaming fashion. The higher the ratio of written vs.\ read data volume, the more relevant the observed behavior becomes.
\revoff
We think that SpecI2M is a valuable extension to the existing WA-avoiding mechanism on x86 CPUs, the non-temporal stores. However, due to lack of insight about the heuristics which govern its activation, building accurate quantitative performance models is difficult. We would prefer if Intel published more detailed information about SpecI2M, which would benefit performance engineers and tool builders alike. 

Preliminary measurements on Intel Sapphire Rapids server CPUs revealed that SpecI2M works but is generally less effective there. In future work, we will carry out a more thorough investigation of SpecI2M and its occasional failure using hardware performance counters on all Intel CPUs that support it. 

\section*{Acknoledgement}
This work was partly funded by BMBF through the DAREXA-F project.

\bibliographystyle{IEEEtran}
\bibliography{IEEEabrv,clv-pmbs}

\fi

\ifdefined\withartifacts

\newpage
\onecolumn
\pagestyle{empty}
\section*{Artifact Identification}

The article investigates a new hardware feature introduced with the Intel Ice Lake SP microarchitecture that automatically avoids write-allocate traffic for streaming write accesses.
The analysis uses the CloverLeaf benchmark code from the SPEChpc 2021 suite and micro-benchmarks to assess the effectiveness of SpecI2M.

The artifacts contain scripts to make CloverLeaf a stand-alone application without the SPEC benchmarking harness and patches for code changes applied to the CloverLeaf benchmark code for the in-depth analysis.
Moreover, depending on the task, there are scripts that gather data for multiple parameters, commonly thread counts.
All data collecting scripts produce CSV files for further analysis or plotting. For reference, the CSV files with the results shown in the paper are part of the artifact description.
Moreover, the artifact description contains the TeX sources to plot all figures presented in this work. They contain the transformations done to the data (scaling from GByte to Byte, calculating ratios, etc.) before plotting. 

With a working installation of the SPEChpc2021 suite, an Intel Ice Lake SP system with a fixed clock frequency of 2.4GHz and applied patches from this artifact description, all benchmarks can be re-run and reproduce the data presented in this work.
It is important to note that the write-allocate avoidance feature behaves differently on different Ice Lake SP chips.
For a full reproduction of the shown results, one has to use systems with Intel Xeon Platinum 8360Y and Intel Xeon Platinum 8480+ as well.
Furthermore, for disabling certain hardware features like the hardware prefetchers, SpecI2M or other undocumented prefetchers, one needs root access to the system.

The tarball containing all files of the artifact description can be downloaded via \url{http://tiny.cc/ipdps24-cloverleaf-tar}.

\section*{Reproducibility of Experiments}
\subsection*{Setup and execution of base CloverLeaf benchmark from SPEChpc 2021}
\begin{enumerate}
\item Install the SPEChpc 2021 suite version 1.1 from original sources.
\item Initialize shell environment for SPEChpc 2021 (\texttt{source shrc}).
\item Load Intel compilers v2021.6.0 and Intel MPI 2021.7.0.
\item Create a configuration file fitting your local system, especially the compilers need to be switched to the Intel legacy compiler \texttt{mpiifort} and the optimization flags to \texttt{OPITMIZE=-g -O3 -xCORE-AVX512 -qopt-zmm-usage=high}. In the following, we use \texttt{MYLOCALCONFIG.cfg} as name for the configuration.
\item Change the \texttt{BASEPATH} variable in \texttt{spec-extract.sh} to point to the SPEChpc2021 installation.
\item Use the \texttt{spec-extract.sh} shell script to get a stand-alone version of the CloverLeaf benchmark with the ``Tiny'' input set: \texttt{./spec-extract.sh -{}-benchmark 519.clvleaf\_t -{}-output \$HOME/PMBS-clvleaf -{}-config MYLOCALCONFIG.cfg -{}-variant ref -{}-tune base}.
\item The stand-alone version of CloverLeaf should be in \texttt{\$HOME/PMBS-clvleaf} and can be compiled using \texttt{make}. Internally, it still uses the SPEC tools \texttt{specmake} and \texttt{specperl}, so the shell environment for SPEChpc2021 is required.
\item The benchmarks are executed on a SLURM-based cluster. All benchmark runs were performed with the base CPU frequency using \texttt{-{}-cpu-freq=2400000-2400000:performance} as parameter for \texttt{sbatch} or as CLI option for \texttt{srun}.
\end{enumerate}

\subsection*{Preparation of the test system}
\begin{enumerate}
\item Activate SNC in the BIOS
\item Activate LLC prefetcher in the BIOS if possible.
\end{enumerate}

\subsection*{Runtime profile using \texttt{gprofng} (Listing~\ref{lst:gprofng})}
\begin{enumerate}
\item The \texttt{gprofng} profiler is part of the \texttt{binutils} package but commonly not installed on systems.
\item Clone the repository: \texttt{git clone https://github.com/oracle/binutils-gdb.git}
\item Configure the build with installation prefix: \texttt{./configure -{}-prefix=\$PREFIX -{}-host-shared-lib}
\item By default, the compilation of \texttt{binutils} does not produce platform-independent code. Consequently, we add \texttt{-fPIC} to all \texttt{CFLAGS} variables in all \texttt{Makefiles}.
\item Build the profiler with \texttt{make}.
\item In the \texttt{gprofng} folder, adjust the variable \texttt{VARIANT} in all \texttt{Makefiles} to \texttt{x86\_64-Linux}.
\item Go to the subfolders \texttt{libcollector} as well as \texttt{src} to compile and install \texttt{make \&\& make install}.
\item Go back to the main \texttt{binutils} folder and install everything with \texttt{make install}.
\item In order to use it, run CloverLeaf wrapped through \texttt{gprofng}: \texttt{srun -{}-cpu-freq=2400000-2400000:performance -N 1 -n 72 -{}-time=01:00:00 gp-collect-app ./clvleaf}. This will create 72 output directories (named \texttt{test.<NUM>.err} with \texttt{NUM} being 1--72.
\item The runtime profile can be read out using \texttt{gprofng display text -script gprofng\_hotspots\_script <PATH/TO/RUNDIR>/test.*.er} with \texttt{gprofng\_hotspots\_script} being a script with the following content (in artifact description):
\begin{minted}[bgcolor=lighter-gray]{text}
metrics name:e.%totalcpu:i.%totalcpu
sort e.totalcpu
limit 11
functions
\end{minted}
\end{enumerate}

\subsection*{Apply patches for code changes}
\begin{enumerate}
\item Apply the patchset 00 in the artifact description tarball for basic changes to the build system.
\item Apply the patchset 01 in the artifact description tarball for all code changes.
\item By default and if not otherwise noted, the options in \texttt{config.mk} are set to:
\begin{itemize}
    \item \texttt{ALIGN\_ARRAYS = ON}
    \item \texttt{LOOP\_BARRIERS = OFF}
    \item \texttt{NT\_STORE\_DIR = OFF}
    \item \texttt{OPTIMIZE\_LOOPS = OFF}
    \item \texttt{HOTSPOTS = OFF}
    \item \texttt{ALL\_HOTSPOT\_LOOPS = OFF}
\end{itemize}
\end{enumerate}

\subsection*{Instrumentation with LIKWID's MarkerAPI and measurements}
\begin{enumerate}
\item Download version 5.2.2 of LIKWID from official sources (Github, FAU FTP). Note that currently, for Sapphire Rapids (SPR) support, one needs to use the \href{https://github.com/RRZE-HPC/likwid/pull/524}{Pull Request \#524} since LIKWID does not support SPR in an official release yet.
\item Change \texttt{FORTRAN\_INTERFACE=true} in \texttt{config.mk}.
\item Load the Intel compilers v2021.6.0 for \texttt{ifort} (which is used by default for LIKWID's Fortran90 interface).
\item Build and install LIKWID with \texttt{make \&\& (sudo) make install}.
\item Activate relevant option (\texttt{HOTSPOTS} or \texttt{ALL\_HOTSPOT\_LOOPS}) in \texttt{config.mk} and rebuild CloverLeaf. Please note that the LIKWID installation on the test system provides environment variables \texttt{LIKWID\_INC=-I\$LIKWID\_INSTALL\_DIR/include} and \texttt{LIKWID\_LIB=-L\$LIKWID\_INSTALL\_DIR/lib} which are used by the build system to find the headers and libraries. If you do not rebuild CloverLeaf, you will get errors from LIKWID: \texttt{Marker API result file does not exist. This may happen if the application has not called LIKWID\_MARKER\_CLOSE}.
\item Use \texttt{likwid-mpirun} to execute and measure: \texttt{for nprocs in `seq 1 72`; do likwid-mpirun -n \$nprocs -m -g MEM\_DP -mpi slurm ./clvleaf; done}. The batch script \texttt{slurm\_job.sh} with the exact SLURM settings is part of the description tarball.
\item If the results in the figures contain error bars, we performed 10 runs to do statistics. See \texttt{slurm\_job.sh} for execution on the test system. It produces output files with the name \texttt{clv.<nprocs>.out.txt}. One needs to evaluate the median and the minimum and maximum values for correctly reproducing the figures with error bars.
\item In order to get the data for all thread counts and all markers, the shell script \texttt{gather\_likwid\_MEMDP.sh} in the artifact description reads all runs and variants and produces a CSV file with the results: \texttt{gather\_likwid\_MEMDP.sh PATH/TO/RUN\_DIR clv.<nprocs> thread\_<nprocs>.dat}
\item In order to run with fixed CPU frequency, you have to add the MPI option to \texttt{likwid-mpirun}: \texttt{likwid-mpirun -{}-mpiopts "-{}-cpu-freq=..."}
\end{enumerate}

\subsection*{Scaling runs (Figure~\ref{fig:speedup})}
\begin{enumerate}
\item LIKWID installation has to be functional.
\item CloverLeaf build has to be functional.
\item Check the \texttt{slurm\_job.sh} file for the right modules loaded (line 25) and remove the \texttt{-m} flag from the \texttt{likwid-mpirun} command in line 48. Now copy it to the CloverLeaf folder.
\item You can run a full scaling run using the script and the following bash command: \texttt{for nprocs in `seq 1 72`; do sbatch -{}-export=ranks=\$nprocs ./slurm\_job.sh; done}
\item After all runs, you will have 72 files in the format \texttt{clv.<nprocs>.out.txt} and one file called \texttt{logs.icx.out}.
\item Use the \texttt{logs.icx.out} file to create a CSV file similar to the \texttt{scaling.csv} file in the artifact description.
\item Run \texttt{./gather\_likwid\_MEMDP.sh PATH/TO/RUN\_DIR clv. clv\_naive\_memdp.csv} to get a CSV file similiar to the \texttt{clv\_naive\_memdp.csv} file in the artifact description tarball.
\item Use the \texttt{scaling.tex} file to create the graph.
\end{enumerate}

\subsection*{Code balance graphs (Figure~\ref{fig:code_balance_naive})}
\begin{enumerate}
\item LIKWID installation has to be functional.
\item CloverLeaf build has to be functional with the following flags set to \texttt{ON} in \texttt{config.mk}: \texttt{ALIGN\_ARRAYS}, \texttt{LOOP\_BARRIERS}, \texttt{ALL\_HOTSPOT\_LOOPS}
\item Check the \texttt{slurm\_job.sh} file for the right modules loaded (line 25) and make sure the \texttt{-m} flag from the \texttt{likwid-mpirun} command in line 47 is set in case you removed it before. Now copy it to the CloverLeaf folder.
\item You can run a full scaling run using the script: \texttt{for nprocs in `seq 1 72`; do sbatch -{}-export=ranks=\$nprocs ./slurm\_job.sh; done}
\item After all runs, you will have 72 files in the format \texttt{clv.<nprocs>.out.txt}.
\item Run \texttt{./gather\_likwid\_MEMDP.sh <PATH/TO/RUN\_DIR> clv. \textbackslash\textbackslash \\ clv\_aligned\_hotspots\_barriers\_memdp.csv} to get a CSV file similiar to the \texttt{clv\_aligned\_hotspots\_barriers\_memdp.csv} file in the artifact description tarball.
\item Use the \texttt{datavol-ac-pdv.tex} and \texttt{datavol-am.tex} files to create the graphs.
\end{enumerate}

\subsection*{MPI measurements using ITAC (Intel Trace Analyzer and Collector) (Figure~\ref{fig:mpi})}
\begin{enumerate}
\item Prepare shell environment to use ITAC, i.e., make sure that the environment variable \texttt{\$VT\_ROOT} as well as \texttt{\$VT\_SLIB\_DIR} is set and that the binary \texttt{traceanalyzer} is available.
\item Rebuild CloverLeaf with ITAC by uncommenting line 38 in the \texttt{Makefile.spec} and run \texttt{make clean \&\& make -j}
\item Uncomment the ITAC-related lines in \texttt{slurm\_job.sh} --- each block marked by \texttt{\#ITAC} and \texttt{\#ITAC-END} --- and comment the original \texttt{likwid-mpirun} command in line 48.
\item Run the bash command \texttt{for nprocs in 2 17 18 19 37 38 71 72; do sbatch -{}-export=ranks=\$nprocs ./slurm\_job.sh; done}.
\item The analysis will be saved as a single \texttt{.stf} file for each run and can be opened and inspected by the Intel Trace Analyzer (\texttt{traceanalyzer}). With the summary of the different reports, create a CSV file similar to the \texttt{mpi.csv} file in the artifact description tarball.
\item Use the \texttt{mpi.tex} file to create the graph.
\end{enumerate}

\subsection*{Micro-benchmarking of \texttt{store} kernel with and without non-temporal stores (Figure~\ref{fig:stores})}
\begin{enumerate}
\item LIKWID installation has to be functional
\item LIKWID contains the tool \texttt{likwid-bench} which provides the \texttt{store} kernel with a single data steam in both variants:
\begin{itemize}
    \item Normal AVX512 stores: \texttt{likwid-bench -t store\_avx512 -W N:10GB:<thread count>}
    \item AVX512 stores with non-temporal hint: \texttt{likwid-bench -t store\_mem\_avx512 -W N:10GB:<thread count>}
\end{itemize}
\item The \texttt{store} kernels with two and three streams are part of the artifact description tarball. Copy the files to the path \newline \texttt{\$HOME/.likwid/bench/x86-64/} to use them like the default \texttt{store} kernels:
\begin{itemize}
\item \texttt{store2\_avx512}: Normal store instructions to two streams
\item \texttt{store3\_avx512}: Normal store instructions to three streams
\item \texttt{store2\_mem\_avx512}: Store instructions with non-temporal hint to two streams
\item \texttt{store3\_mem\_avx512}: Store instructions with non-temporal hint to three streams
\end{itemize}
\item \texttt{likwid-bench} already contains Marker API calls. In order to measure the memory data volume, one can wrap it with \texttt{likwid-perfctr}: \texttt{likwid-perfctr -C <cpu selection> -g MEM -m likwid-bench -t <kernel> -W N:10GB:<thread count>}
\item In order to get the data for all thread counts, the shell script \texttt{store\_ubenchs.sh <kernel>} in the artifact description executes the benchmark with all thread counts and produces a CSV file with the results.
\item The experiment needs to be repeated (at least) 10 times to determine the minimum, maximum and median value in the figure.
\item Create CSV files similar to \texttt{st1.csv}, \texttt{st2.csv}, \texttt{st3.csv}, \texttt{stnt1.csv}, \texttt{stnt2.csv}, \texttt{stnt3.csv} in the artifact description tarball.
\item Use the \texttt{stores.tex} file to create the graph.
\end{enumerate}

\subsection*{Hardware performance counter measurements of \texttt{copy} kernel (Figure~\ref{fig:copy_speci2m})}
\begin{enumerate}
\item LIKWID installation has to be functional
\item LIKWID contains the tool \texttt{likwid-bench} which provides the \texttt{copy\_avx} kernel
\item Create a file \texttt{\$HOME/.likwid/groups/ICX/SPECI2M.txt} with the content in Listing~\ref{lst:likwid_speci2m_group}
\item Run and measure benchmark:
\begin{minted}[bgcolor=lighter-gray]{text}
for thr in {1..72}; do
  likwid-perfctr -C E:N:${thr} -g SPECI2M -m likwid-bench -t copy_avx -W N:2GB
done
\end{minted}
If the test system has SMT activated, use \texttt{E:N:\$\{thr\}:1:2}.
\item Record values of
\begin{itemize}
    \item \texttt{\$\{thr\}}
    \item \texttt{Memory read data volume [GBytes]}
    \item \texttt{Memory write data volume [GBytes]}
    \item \texttt{SpecI2M data volume [GBytes]}
    \item \texttt{Iterations per thread}
    \item \texttt{Inner loop executions}
\end{itemize}
\item For calculating the data volume per element in Bytes, one has to derive: \texttt{\$\{thr\}}$\times$\texttt{Iterations per thread}$\times$\texttt{Inner loop executions}$\times$16. The last factor is related to the \texttt{copy\_avx} kernel which processes 16 elements per inner loop iteration (2 cache lines)
\end{enumerate}

\begin{minted}[
frame=lines,
framesep=2mm,
baselinestretch=1.2,
bgcolor=lighter-gray,
fontsize=\footnotesize,
]{text}
SHORT  Memory bandwidth in MBytes/s including SpecI2M

EVENTSET
FIXC0 INSTR_RETIRED_ANY
FIXC1 CPU_CLK_UNHALTED_CORE
FIXC2 CPU_CLK_UNHALTED_REF
CBOX0C0 TOR_INSERTS_IA_ITOM
CBOX1C0 TOR_INSERTS_IA_ITOM
CBOX2C0 TOR_INSERTS_IA_ITOM
CBOX3C0 TOR_INSERTS_IA_ITOM
CBOX4C0 TOR_INSERTS_IA_ITOM
CBOX5C0 TOR_INSERTS_IA_ITOM
CBOX6C0 TOR_INSERTS_IA_ITOM
CBOX7C0 TOR_INSERTS_IA_ITOM
CBOX8C0 TOR_INSERTS_IA_ITOM
CBOX9C0 TOR_INSERTS_IA_ITOM
CBOX10C0 TOR_INSERTS_IA_ITOM
CBOX11C0 TOR_INSERTS_IA_ITOM
CBOX12C0 TOR_INSERTS_IA_ITOM
CBOX13C0 TOR_INSERTS_IA_ITOM
CBOX14C0 TOR_INSERTS_IA_ITOM
CBOX15C0 TOR_INSERTS_IA_ITOM
CBOX16C0 TOR_INSERTS_IA_ITOM
CBOX17C0 TOR_INSERTS_IA_ITOM
CBOX18C0 TOR_INSERTS_IA_ITOM
CBOX19C0 TOR_INSERTS_IA_ITOM
CBOX20C0 TOR_INSERTS_IA_ITOM
CBOX21C0 TOR_INSERTS_IA_ITOM
CBOX22C0 TOR_INSERTS_IA_ITOM
CBOX23C0 TOR_INSERTS_IA_ITOM
CBOX24C0 TOR_INSERTS_IA_ITOM
CBOX25C0 TOR_INSERTS_IA_ITOM
CBOX26C0 TOR_INSERTS_IA_ITOM
CBOX27C0 TOR_INSERTS_IA_ITOM
CBOX28C0 TOR_INSERTS_IA_ITOM
CBOX29C0 TOR_INSERTS_IA_ITOM
CBOX30C0 TOR_INSERTS_IA_ITOM
CBOX31C0 TOR_INSERTS_IA_ITOM
CBOX32C0 TOR_INSERTS_IA_ITOM
CBOX33C0 TOR_INSERTS_IA_ITOM
CBOX34C0 TOR_INSERTS_IA_ITOM
CBOX35C0 TOR_INSERTS_IA_ITOM
CBOX36C0 TOR_INSERTS_IA_ITOM
CBOX37C0 TOR_INSERTS_IA_ITOM
CBOX38C0 TOR_INSERTS_IA_ITOM
CBOX39C0 TOR_INSERTS_IA_ITOM
MBOX0C0 CAS_COUNT_RD
MBOX0C1 CAS_COUNT_WR
MBOX1C0 CAS_COUNT_RD
MBOX1C1 CAS_COUNT_WR
MBOX2C0 CAS_COUNT_RD
MBOX2C1 CAS_COUNT_WR
MBOX3C0 CAS_COUNT_RD
MBOX3C1 CAS_COUNT_WR
MBOX4C0 CAS_COUNT_RD
MBOX4C1 CAS_COUNT_WR
MBOX5C0 CAS_COUNT_RD
MBOX5C1 CAS_COUNT_WR
MBOX6C0 CAS_COUNT_RD
MBOX6C1 CAS_COUNT_WR
MBOX7C0 CAS_COUNT_RD
MBOX7C1 CAS_COUNT_WR

METRICS
Runtime (RDTSC) [s] time
SpecI2M data volume [GBytes] 1.0E-09*(CBOX0C0+CBOX1C0+CBOX2C0+CBOX3C0+CBOX4C0+CBOX5C0+CBOX6C0+CBOX7C0 \\
  +CBOX8C0+CBOX9C0+CBOX10C0+CBOX11C0+CBOX12C0+CBOX13C0+CBOX14C0 \\
  +CBOX15C0+CBOX16C0+CBOX17C0+CBOX18C0+CBOX19C0+CBOX20C0+CBOX21C0 \\
  +CBOX22C0+CBOX23C0+CBOX24C0+CBOX25C0+CBOX26C0+CBOX27C0+CBOX28C0 \\
  +CBOX29C0+CBOX30C0+CBOX31C0+CBOX32C0+CBOX33C0+CBOX34C0+CBOX35C0 \\
  +CBOX36C0+CBOX37C0+CBOX38C0+CBOX39C0)*64.0
Memory read data volume [GBytes] 1.0E-09*(MBOX0C0+MBOX1C0+MBOX2C0+MBOX3C0+MBOX4C0+MBOX5C0+MBOX6C0 \\
  +MBOX7C0)*64.0
Memory write data volume [GBytes] 1.0E-09*(MBOX0C1+MBOX1C1+MBOX2C1+MBOX3C1+MBOX4C1+MBOX5C1+MBOX6C1 \\
  +MBOX7C1)*64.0

LONG
Formulas:
\end{minted}
\captionof{listing}{Performance group for LIKWID usable for Intel ``Ice Lake SP'' to count SpecI2M and memory data traffic}
\label{lst:likwid_speci2m_group}

\subsection*{Phenomenological Model (Figure~\ref{fig:perf-model})}
\begin{enumerate}
\item LIKWID installation has to be functional.
\item For the ``Original measurement'' data series, you can use the measurements for 72 cores from Figure 2.
\item For the ``Optimized measurement'' data series, the CloverLeaf build has to be functional with the following flags set to \texttt{ON} in \texttt{config.mk}: \texttt{ALIGN\_ARRAYS}, \texttt{LOOP\_BARRIERS}, \texttt{ALL\_HOTSPOT\_LOOPS}, \texttt{NT\_STORE\_DIR}, \texttt{OPTIMIZE\_LOOPS}
\item Check the \texttt{slurm\_job.sh} file for the right modules loaded (line 25) and make sure the \texttt{-m} flag from the \texttt{likwid-mpirun} command in line 48 is set in case you removed it before. Now copy it to the CloverLeaf folder.
\item You can run a full scaling run using the script: \texttt{sbatch -{}-export=ranks=72 ./slurm\_job.sh}
\item Use the \texttt{model.csv} from the artifact description tarball and update the column \texttt{72c\_meas} and \texttt{opt\_meas} with the values from your original measurement and optimized measurement, respectively.
\item Use the \texttt{loops-prediction.tex} file to create the graph.
\end{enumerate}

\subsection*{Disabling CPU prefetchers}
\begin{enumerate}
\item LIKWID installation has to be functional
\item Adjust CPU prefetchers:
\begin{itemize}
    \item Enable all prefetchers: \newline
    \texttt{likwid-features -c N -e HW\_PREFETCHER,CL\_PREFETCHER,DCU\_PREFETCHER,IP\_PREFETCHER}
    \item Disable all prefetchers: \newline
    \texttt{likwid-features -c N -d HW\_PREFETCHER,CL\_PREFETCHER,DCU\_PREFETCHER,IP\_PREFETCHER}
\end{itemize}
\item The information on how to disable SpecI2M and further undocumented prefetchers on a system is under NDA and must be requested from Intel.
\end{enumerate}

\subsection*{Read-to-Write Data Volume of the Copy Benchmark (Figure~\ref{fig:copy})}
\begin{enumerate}
\item LIKWID installation has to be functional.
\item Download \texttt{TheBandwidthBenchmark} from \url{https://github.com/RRZE-HPC/TheBandwidthBenchmark} and apply the patch in the artifact description tarball. You have to have the Intel legacy compiler ICC in your software environment for compilation.
\item Run the following bash command to create 54 files in the format \texttt{copy.<dim>.<halo>.out.txt}:
\begin{minted}[bgcolor=lighter-gray]{text}
for dim in 216 530 1920; do
  sed -r -i "s/inner_size = [0-9]+/inner_size = $dim/" src/copy.c
  for h in `seq 0 17`; do
    sed -r -i "s/halo = [0-9]+/halo = $h/" src/copy.c
    make 
    likwid-perfctr -C E:N:72 -g MEM_DP -m ./bwbench-ICC > copy.$dim.$h.out.txt
  done
done
\end{minted}
\item Run the \texttt{gather\_likwid\_MEMDP\_copy.sh} in the following way to create the CSV files similiar to \texttt{copy216.csv}, \texttt{copy530.csv}, and \texttt{copy1920.csv} in the artifact description tarball: \texttt{for dim in 216 530 1920; do ./gather\_likwid\_MEMDP\_copy.sh /PATH/TO/BWBENCH/OUTPUT/ copy.\$dim. \$dim copy\$dim.csv; done}
\item Disable all CPU prefetchers as described in the Section above and repeat the last steps to create the CSV files \newline \texttt{copy216-pfoff.csv}, \texttt{copy530-pfoff.csv}, and \texttt{copy1920-pfoff.csv}. Make sure you (re)moved the previous output files before.
\item Use the \texttt{stores.tex} file to create the graph.
\end{enumerate}

\subsection*{Estimated Execution Time of the Experiment Workflow}
A single-core CloverLeaf run with the ``Tiny'' data set at a fixed CPU clock frequency of 2.4 GHz should take approximately 2.5 hours.
Depending on the number of available nodes, one can execute all CloverLeaf benchmarks within this amount of time in parallel.
If executed one by one, the worst-case scenario amounts to an execution time of 28.5 hours.

\texttt{likwid-bench} runs each kernel at least for 1 seconds and minimally 10 iterations over the data set which can result in a runtime of around 10 seconds per run, especially for runs with a low thread count and large data set.

Measurements of a whole application run with \texttt{likwid-perfctr} do not add any overhead during execution. The MarkerAPI instrumentation adds measurable overhead to the application run if the markers are called frequently. Besides the marker code itself, the accesses to the hardware registers during the runtime of the application accounts for most of the runtime overhead.

Note that an exclusive usage of the compute node is necessary for accurate results and measurements due to the excessive usage of memory bandwidth.

\subsection*{Description of the expected results and relation to article}
\begin{enumerate}
\item For the initial scaling runs with LIKWID measurements, one should see increasing data bandwidth within a NUMA domain and linear progression afterwards. There will be drops in memory bandwidths if the number of processes is prime. Moreover, one should see inside the first NUMA domain that despite the bandwidth saturates, the performance/speedup keeps increasing due to WA evasion getting more effective near saturation.
\item For the runtime profile, one should expect the three functions \texttt{advec\_mom\_kernel}, \texttt{advec\_cell\_kernel} and \texttt{pdv\_kernel} to be the top three functions.
\item For the required data volume per iteration, one has to divide the ``Memory data volume'' (in Bytes) reported by LIKWID by the ``call count'', the amount of instrumentation traversals for the specific multipled by the timesteps (400) and the grid size ($15360^2$). For most kernels, the data volume per iteration should decrease in the first NUMA domain. Afterwards the volume per iteration is almost stable with upward spikes when the process count is prime.
\item When opening the \texttt{.stf} file with the Intel Trace Analyzer, one should see a distribution similar to Figure 3, i.e., a serial execution share of 94--99\% with \texttt{MPI\_Waitall} and \texttt{MPI\_Allreduce} taking two thirds and one third of the overall MPI runtime, respectively. The runs with 19, 37, 38, and 71 cores should show at least twice as much of the relative runtime than the runs with 2, 17, 18, and 72 cores.
\item For the store ratio tests, one has to run \texttt{likwid-bench} wrapped by \texttt{likwid-perfctr -g MEM} and divide the ``Memory data volume'' returned by \texttt{likwid-perfctr} by the data volume reported by \texttt{likwid-bench}. The ratios should decrease for the kernels with normal stores in the first NUMA domain. Afterwards the ratio should rise again to fall back down at the end of the second NUMA domain. This behavior, to a smaller extent, is also visible on the following NUMA domains. For the non-temporal stores, the ratio for a single, two or three streams should be almost identical. Although all write-allocates should be avoided, the ratio should increase slightly from 1.0 to 1.15.
\item For the phenomenological model, one should get measurements within the range of 2\% compared to the model for the optimized measurements, while the original measurements are slightly worse (in average 6\%), with a specifically high deviation from the optimized measurements for the loops ac02 and ac05.
\item For the copy load-to-store ratio benchmark, one has to run TheBandwidthBenchmark with applied code patches wrapped by \texttt{likwid-perfctr} with the \texttt{MEM\_DP} group. To get the ratio, one has to divide the total read data volume by the total write data volume. The larger the inner dimension is, the lower the  load-to-store ratio should be. One should see a significant reduction of the load-to-store ratio for HALO sizes creating full cache lines after one, two, or four iterations for the cases 216 (with and without prefetchers) and 1920 (without prefetchers). 
\end{enumerate}
\fi

\end{document}